\documentclass{aa}  

\usepackage{graphicx}
\usepackage{txfonts}
\usepackage{arydshln}
\usepackage{booktabs}
\usepackage{amsfonts}

\newcommand{\yemin}{$Y_\mathrm{e}^\mathrm{min}$}
\newcommand{\yedot}{$\dot{Y}_\mathrm{e}$}
\begin{document}

   \title{Do electron-capture supernovae make neutron stars?}

   \subtitle{First multidimensional hydrodynamic simulations of the oxygen
deflagration}

   \author{S. Jones
          \inst{1}\fnmsep\thanks{Alexander von Humboldt Fellow}
          \and
          F. K. R\"{o}pke\inst{1}\fnmsep\inst{2}
          \and
          R. Pakmor\inst{1}
          \and
          I. R. Seitenzahl\inst{3}\fnmsep\inst{4}
          \and
          S. T. Ohlmann\inst{1}\fnmsep\inst{2}
          \and
          P. V. F. Edelmann\inst{1}
          }

   \institute{Heidelberg Institute for Theoretical Studies,
              Schloss-Wolfsbrunnenweg 35, 69118 Heidelberg, Germany\\
              \email{samuel.jones@h-its.org}
         \and
             Zentrum für Astronomie der Universität Heidelberg,
             Albert-Ueberle-Str. 2, D-69120 Heidelberg, Germany
         \and
             Research School of Astronomy and Astrophysics, Australian
             National University, Canberra, ACT 2611, Australia
          \and
             ARC Centre of Excellence for All-Sky Astrophysics (CAASTRO)
             }

   \date{Received 17 February 2016; accepted 4 July 2016}

 
  \abstract
   {In the classical picture, electron-capture supernovae and the
accretion-induced collapse of oxygen-neon white dwarfs undergo an oxygen
deflagration phase before gravitational collapse produces a neutron star. These
types of core collapse events are postulated to explain several astronomical
phenomena.  In this work, the oxygen deflagration phase is simulated for the
first time using multidimensional hydrodynamics.}
   {By simulating the oxygen deflagration with multidimensional hydrodynamics
and a level-set-based flame approach, new insights can be gained into the
explosive deaths of $8-10~M_\odot$ stars and oxygen-neon white dwarfs that
accrete material from a binary companion star. The main aim is to determine
whether these events are thermonuclear or core-collapse supernova explosions,
and hence whether neutron stars are formed by such phenomena.} 
   {The oxygen deflagration is simulated in oxygen-neon cores with three
different central ignition densities. The intermediate density case is perhaps
the most realistic, being based on recent nuclear physics calculations and 1D
stellar models. The 3D hydrodynamic simulations presented in this work begin
from a centrally confined flame structure using a level-set-based flame
approach and are performed in $256^3$ and $512^3$ numerical resolutions.}
   {In the simulations with intermediate and low ignition density, the cores do
not appear to collapse into neutron stars. Instead, almost a solar mass of
material becomes unbound from the cores, leaving bound remnants. These
simulations represent the case in which semiconvective mixing during the
electron-capture phase preceding the deflagration is inefficient. The masses
of the bound remnants double when Coulomb corrections are included in the
equation of state, however they still do not exceed the effective Chandrasekhar
mass and, hence, would not collapse into neutron stars. The simulations with
the highest ignition density ($\log_{10}\rho_{\rm c}=10.3$), representing the
case where semiconvective mixing is very efficient, show clear signs that the
core will collapse into a neutron star.}
  {}

   \keywords{stars: evolution -- 
             stars: interiors --
             stars: neutron --
             supernovae: general
               }

   \maketitle
%

\section{Introduction}

Electron-capture supernovae (ECSNe) and the accretion-induced collapse
of oxygen-neon (ONe) white dwarfs (WDs) are phenomena in which a degenerate ONe
core is postulated to collapse into a neutron star
\citep{Miyaji1980,Nomoto1984,Nomoto1987,Nomoto1991,Kitaura2006,Fischer2010,
Jones2013,Takahashi2013,Schwab2015}. In the former case, if the SN is from a
single star, a type IIP or IIn-P supernova would be produced with a luminosity
lower than \citep{Smith2013crab} or similar to \citep{Tominaga2013,Moriya2014}
standard type~IIP SNe. It is likely that the progenitor stars of ECSNe have a
binary companion \citep{Sana2012,Dunstall2015}, making it also possible for
ECSNe to produce SNe of type Ib/c, depending upon the degree of stripping by
the companion star \citep{Tauris2013}.

The collapse of the ONe core is triggered by electron captures on
$^{24}\mathrm{Mg}$ and, more importantly, $^{20}\mathrm{Ne}$, which releases
enough energy through the $\gamma$-decay of $^{20}\mathrm{O}$ to ignite
O-burning via $^{16}\mathrm{O}+{^{16}\mathrm{O}}$ fusion
\citep{Miyaji1980,Nomoto1984,Nomoto1987}. Consequently, and owing to the high
degree of degeneracy of the material, a thermonuclear runaway ensues since
there is little to no expansion resulting from the temperature increase.
Instead, the increase in temperature just accelerates the rate of fusion until
the temperature becomes so high (approximately $10^{10}$~K) that the
composition can be said to be in nuclear statistical equilibrium (NSE;
\citealp[e.g.][]{Seitenzahl2009}).  At such temperatures the material is no
longer completely degenerate and can expand in response to the nuclear binding
energy that has been released at the burning front.

\begin{figure*}
\centering
\includegraphics[width=\linewidth]{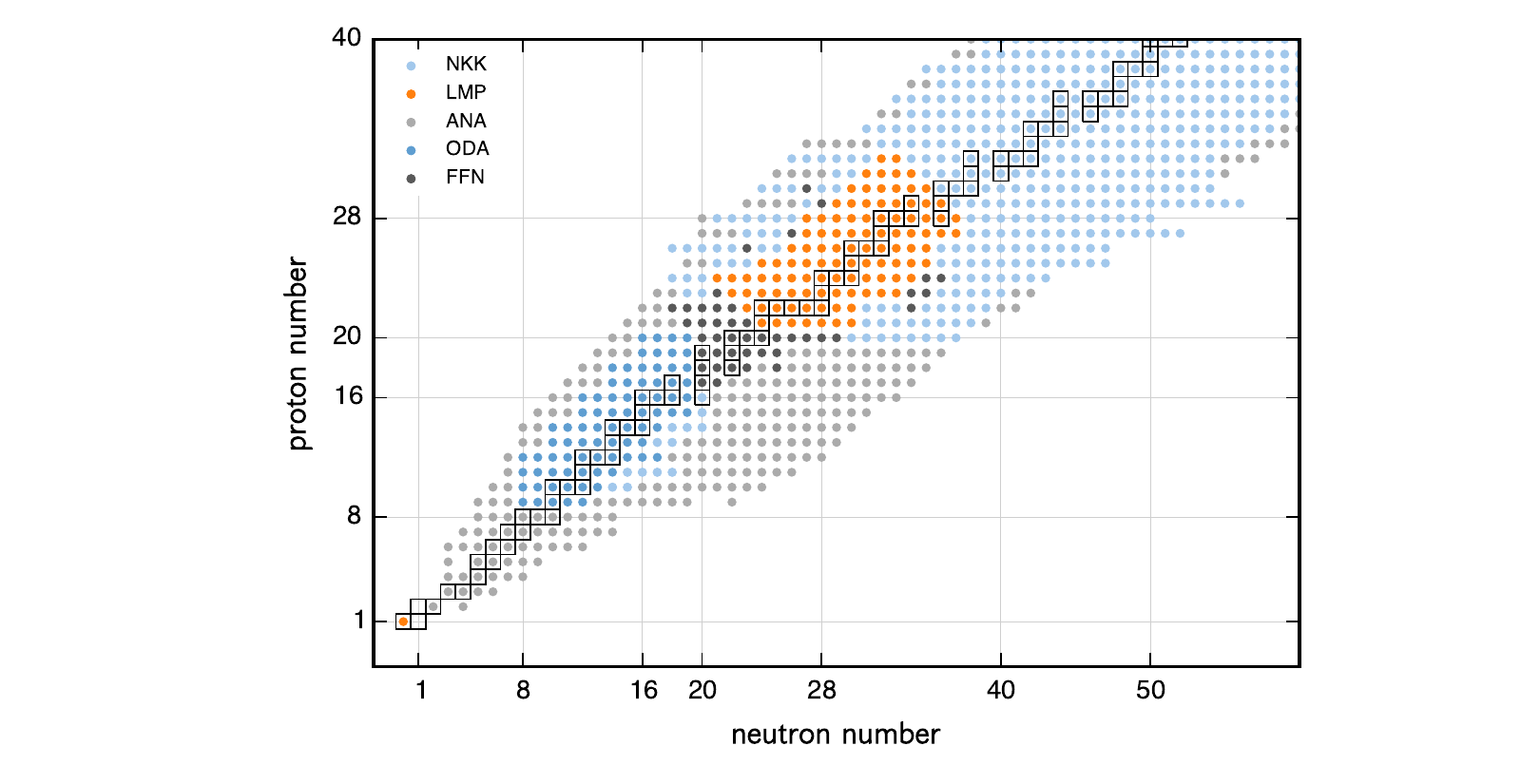}

\caption{Sources of electron-capture rates considered for the calculation of
$dY_\mathrm{e}/dt$ in the NSE ashes of the O deflagration. The inverse
($\beta$-decay) rates were taken from consistent sources. The labels are as
follows: \citet[][NKK]{Nabi2004}, \citet[][LMP]{LMP2001}, \citet[][ODA]{ODA94}
and \citet[][FFN]{FFNweak1985}. ANA corresponds to nuclei for which there are
no tabulated rates available and thus we use an analytical formula similar to
that used by \citet{Arcones2010} and \citet{Sullivan2016}.}

\label{fig:rate-sources}
\end{figure*}

The conservation laws of hydrodynamics allow for two distinct modes of the
propagation of combustion fronts, if modelled as discontinuities between fuel
and ash: a subsonic deflagration where the flame is mediated by thermal
conduction and a supersonic detonation driven by shock waves. Subsonic burning
fronts are subject to a variety of hydrodynamical instabilities and produce
turbulence. The interaction with turbulent eddies increases the flame surface
area and accelerates the flame propagation significantly (see
\citealp{Roepke2009} for a review).  The deflagration model of combustion is
the one most commonly assumed to describe the nuclear burning resulting from
electron capture by $^{20}$Ne and the ignition of $^{16}$O fusion in degenerate
ONe cores, and in the present work this assumption is held.  Assuming that the
front propagates as a deflagration, at the high densities in the ONe core the
flame width is of sub-centimetre order \citep{Timmes1992} and propagates either
by electron conduction (laminar regime) or, if the flame enters the turbulent
burning regime, by turbulence.

Models of stars that develop degenerate ONe cores -- the so-called super-AGB
stars \citep[see, e.g.,][]{Siess2007,Siess2010} -- are riddled with
uncertainties, primarily concerning mass loss \citep{Poelarends2008} and
convective boundary mixing \citep{Denissenkov2013cflame,Jones2016}. If
super-AGB stars eject their envelope quiescently by winds or in a dynamical
event \citep[e.g.,][]{Lau2012} before the ONe core can grow to the
Chandrasekhar mass $M_\mathrm{Ch}$ then an ONe WD and planetary nebula will be
formed. It is still possible that the ONe WD will reach $M_\mathrm{Ch}$ by mass
accretion from a binary companion. If this occurs and the material is retained
by the WD, an O deflagration is ignited owing to electron captures on
$^{20}$Ne, the interior evolution proceeds in much the same way as is described
above for ECSNe \citep[e.g.,][]{Schwab2015} and the star is thought to collapse
into a neutron star \citep{Nomoto1991}. This is known as accretion-induced
collapse (AIC).

The ignition density of the O deflagration has been known to be in the region
of $10^{10}$~g~cm$^{-3}$ for decades \citep{Miyaji1980,Nomoto1987}, although
its precise value is still a topic of much discussion. Several studies
involving the improvement of weak reaction rates for $sd$-shell nuclei have
been undertaken
\citep{Takahara89,Toki2013,Lam2014,Martinez-Pinedo2014,Schwab2015} since the
seminal works of \citet{FFN80,FFN1982a,FFNweak1985} and \citet{ODA94}.  Most
recently, efforts by \citet{Martinez-Pinedo2014} and \citet{Schwab2015}, with
an analytic approach to $^{20}\mathrm{Ne}$ electron capture including a
potentially crucial second-forbidden transition, predict deflagration ignition
densities between about $\log_{10}\rho=9.9$ and $9.95$. 

At such high densities, the NSE ashes deleptonise rapidly ($dY_\mathrm{e}/dt$
strongly negative). The rapid removal of the electrons reduces their
contribution to the pressure, whose gradient supports the star against its own
gravity. The result of removing electrons from a completely degenerate core is
contraction. Contraction increases the density and along with it, the
electron-capture rates. A runaway process ensues thusly and the core of the
star, in the current wisdom \citep{Jones2013,Takahashi2013,Schwab2015}, is
believed to collapse rapidly to a neutron star (more rapidly than a typical
iron-core progenitor), launching a shock wave and producing a dim supernova
\citep{Kitaura2006,Fischer2010} with a relatively low explosion energy and low
Ni ejecta mass.

The promptness of the supernova explosion in this case is caused by the steep
density gradient at the edge of the core in the progenitor stars of ECSNe and
the AIC of ONe WDs. In this more rapid explosion there is postulated to be less
time for asymmetries to develop that would have given the newly-born neutron
star a natal kick. This distinction between ECSNe/AIC and more massive Fe-core
supernovae has been proposed by \citet{Knigge2011} to be responsible for the
observed bimodality in the orbital eccentricities of BeX systems. However, the
density profiles of the lowest mass Fe-core collapse progenitor stars are very
similar indeed to those of ECSN progenitor stars \citep{Jones2013,Woosley2015}.

\begin{figure}
\centering
\includegraphics[width=1.\linewidth,trim={0mm 6mm 0mm 0mm}]{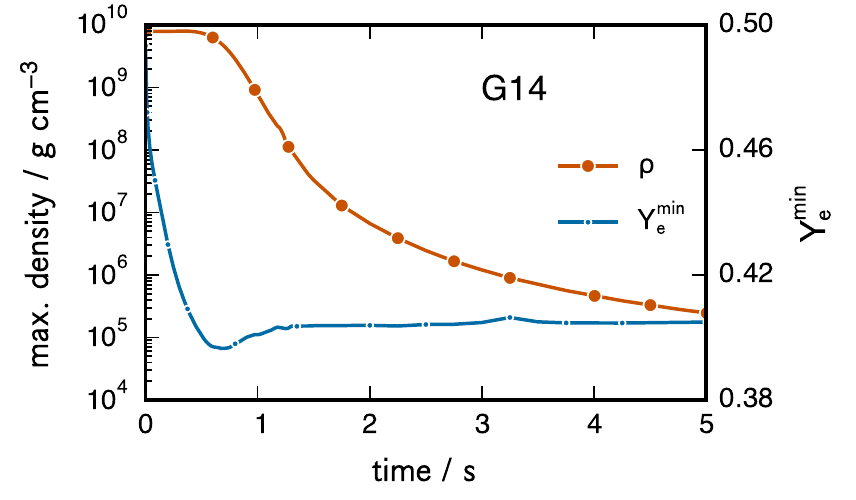} \\
\includegraphics[width=1.\linewidth,trim={0mm 6mm 0mm 0mm}]{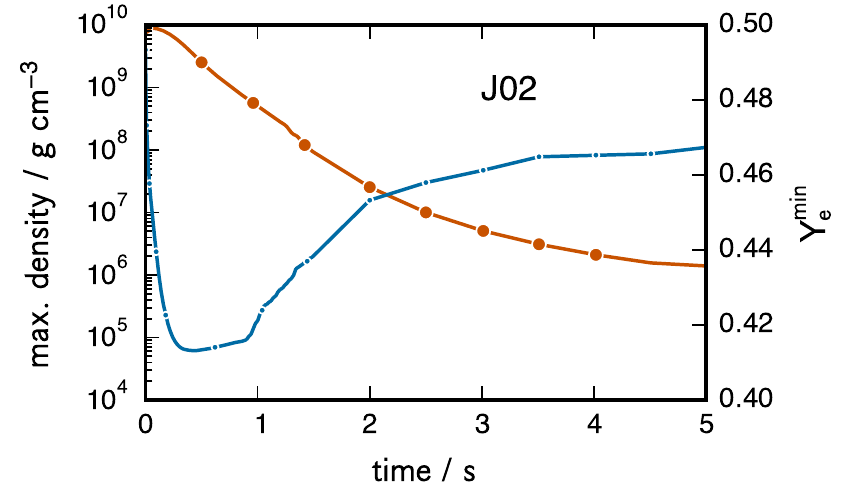} \\
\includegraphics[width=1.\linewidth]{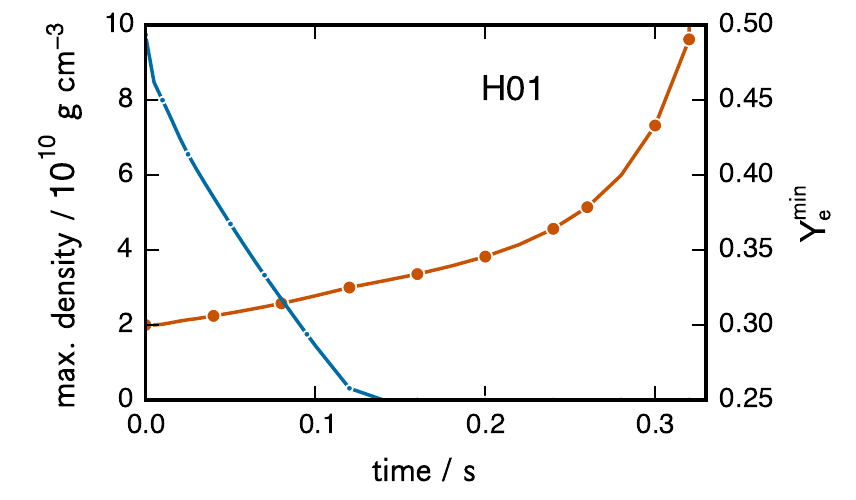}

\caption{Maximum density and minimum electron fraction $Y_\mathrm{e}$ in the
simulations G14, J02 and H01 (see Table~\ref{tab:model-props}). In the G and J
simulations ($\log_{10}\rho_\mathrm{c}^\mathrm{ign}=9.9$ and 9.95,
respectively, the maximum density drops by several orders of magnitude in the
first 5 seconds despite the marked decrease in the minimum $Y_\mathrm{e}$,
leading to the partial disruption of the core and the formation of an ONeFe
white dwarf that does not collapse to form a neutron star. In the H01
simulation ($\log_{10}\rho_\mathrm{c}^\mathrm{ign}=10.3$), the maximum density
only increases with time, reaching $10^{11}$~g~cm$^{-3}$ in the first
$\sim330$~ms.  The simulation was not continued beyond this point because
neutrino interactions with matter were not included in the microphysics,
however the most likely outcome is collapse into a neutron star.}

\label{fig:rhomax-yemin-t}
\end{figure}

In reality the outcome of ECSNe and the AIC of ONe WDs is not such a closed
case. The star's fate balances on a knife edge between collapsing into a
neutron star or becoming completely unbound by the energy released in the
thermonuclear runaway as the deflagration runs through the star. It is a
situation in which the ignition density of the deflagration, the growth of the
Rayleigh--Taylor (R--T) instability, the speed of the deflagration in both the
laminar and turbulent regimes, and both electron capture and $\beta$-decay of
the NSE composition all play critical roles.

Perhaps the most important works highlighting and addressing some of the
questions surrounding the O deflagration phase in ECSNe and the AIC of ONe WDs
are \citet{Nomoto1991}, \citet{Isern1991}, \citet{Canal1992} and
\citet{Timmes1992}. \citet{Nomoto1991} showed that for an ONe core with a
central ignition density of $9.95\times10^9$~g cm$^{-3}$, the distinction
between core collapse and thermonuclear explosion lay with the speed of the
convective or turbulent deflagration wave, for which a time-dependent
formulation of mixing length theory (MLT) was used in their 1D models. For the
mixing length parameter $\alpha=0.7$, which reproduces well the observables of
type Ia supernovae when used in carbon deflagration simulations
\citep{NomotoThielemann1984}, Nomoto \& Kondo found that O deflagration results
in core collapse, while for $\alpha=1.0$ and 1.4 it results in a thermonuclear
explosion (their Fig.~1). 

\citet{Isern1991} also reported that if the O deflagration were to enter the
turbulent burning regime, it could result in either the complete disruption of
the ONe core/WD or in the partial ejection of material, leaving behind a bound
remnant composed of O, Ne and Fe-group elements. Isern et al. used the
parameterisation suggested by \citet{Woosley1986} for the speed of a turbulent
flame as the propagation speed of the O deflagration, with the two free
parameters set to the estimates given by \citet{Woosley1986} together with the
formula. An update to this work by \citet{Canal1992} used the better-resolved
electron-capture rate tables by \citet{Takahara89} and the Ledoux criterion for
convection, finding the ignition density of the deflagration to be lower than
previously thought. The authors found that even if the flame were to remain in
the laminar burning regime, for such a low ignition density
($\sim8.5\times10^9$~g~cm$^{-3}$) the star would become completely unbound.

The simulations of \citet{Canal1992} did not include Coulomb corrections to the
electron-capture reaction rates, which were later shown by \citet[][but see the
footnote in that paper with regards to Ramon Canal's own
findings]{Hashimoto1993} to bring the ignition density back up to
$9\times10^9$~g cm$^{-3}$. Later still, \citet{Gutierrez1996} found an ignition
density of $9.7\times10^9$~g cm$^{-3}$, and although the authors included
Coulomb effects in the electron-capture reaction rates, they opted to use rates
that were more up-to date \citep{ODA94} over those that were better resolved
\citep{Takahara89}, which actually may have resulted in a less accurate
calculation owing to the undersampling issues of the \citet{ODA94} rates for
these conditions (illustrated in Fig.~6 of \citealp{Jones2013} and Fig.~34 of
\citealp{MESA2015}). Another difference between the work of
\citet{Gutierrez1996} and preceding works was the composition of the ONe core.
Where previously $^{20}$Ne had been considered the most abundant nuclear
species based on the models of \citet{Miyaji1987}, Gutierrez et al. assumed an
oxygen-dominated composition based on the models of \citet{Dominguez1993}.
Interestingly, the ignition density found by \citet{Canal1992} is actually in
rather good agreement with very recent predictions by \citet{Schwab2015}, who
used the capture rates of \citet{Martinez-Pinedo2014} \emph{including} Coulomb
effects and also assumed the Ledoux criterion for determining convective
stability.

The calculations by \citet{Timmes1992} of laminar flame speeds and linear
analysis of the growth rate of the R--T instability demonstrates just how
marginal the situation really is, motivating multidimensional hydrodynamic
simulations of the O deflagration phase.  In this paper, we report on the
progress of such 3D hydrodynamic simulations of the O deflagration that we have
performed. Section~\ref{sec:methods} is a description of the methodology and in
Section~\ref{sec:results} the results of the simulations are presented.
Finally, Section~\ref{sec:summary} is a summary of the present work, including
a discussion of the uncertainties.


\section{Methodology}
\label{sec:methods}

Degenerate cores comprised homogeneously of a 65\%/35\% mix of
$^{16}\mathrm{O}$ and $^{20}\mathrm{Ne}$ are prescribed a central density and
integrated into a hydrostatic equilibrium configuration.  They are assumed to
be isothermal, with a temperature of $5\times10^5$~K.  In the evolution
preceding the formation of an ONe core with this central density (in
particular during C burning, the Urca process phase and electron capture on
$^{24}\mathrm{Mg}$), the electron fraction $Y_\mathrm{e}$ would have decreased
slightly from 0.5 (i.e., all nuclei have $N=Z$). This is important for
correctly evaluating the density, and hence the pressure, during the setup.
This is accounted for by assuming a constant $Y_\mathrm{e}=0.493$ for the whole
core, whose value is a mass-weighted average of the $8.75~M_\odot$ progenitor
model from \citet{Jones2013}, which was computed with the MESA stellar
evolution code \citep{MESA2011,MESA2013,MESA2015}. If the $8.8~M_\odot$ model
of \citet{Jones2013} had been considered, in which an episode of off-centre Ne
and O shell burning had taken place prior to the ignition of the O
deflagration, the average $Y_\mathrm{e}$ would have been 0.490 due to the
production of neutron-rich Si-group nuclei. This is a special case that is in
itself interesting for a potentially small fraction of ECSN progenitor stars,
although its realisation depends upon the uncertain physics of flame quenching
in degenerate cores by convective boundary mixing
\citep{Denissenkov2013cflame,Jones2014,Farmer2015,Lecoanet2016}.

\subsection{3D hydrodynamics and the level-set based flame}

\begin{figure}
\centering
\includegraphics[width=1.\linewidth]{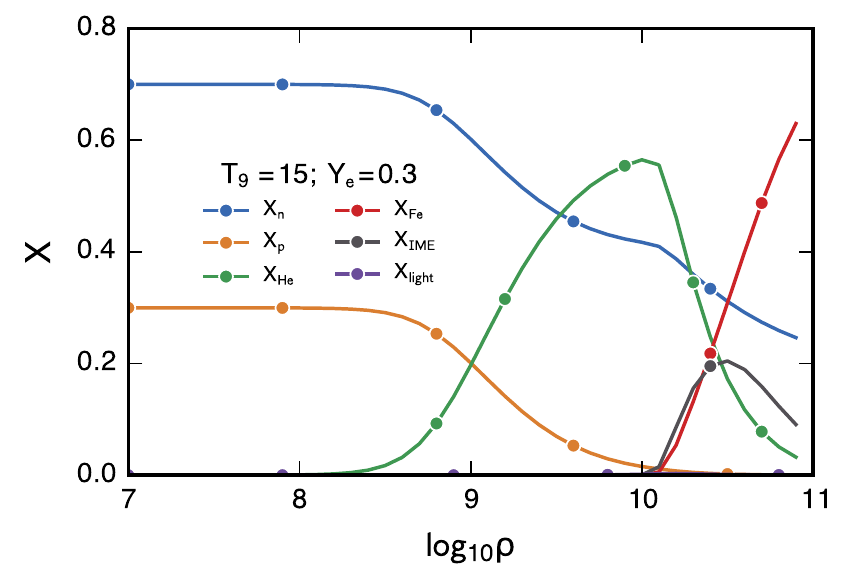}

\caption{NSE composition as a function of density for the conditions
encountered in the simulations with the highest ignition densities
($2\times10^{10}$~g~cm$^{-3}$; $T=1.5\times10^{10}$~K and $Y_\mathrm{e}=0.3$).
Fe denotes elements with $Z\ge24$ and IME the intermediate-mass elements with
$13\le Z<24$. The residual is the light elements excluding p, n and
$\alpha$-particles.}

\label{fig:nse-abunds}
\end{figure}

The O deflagration is followed in 3D using the LEAFS code
\citep{Reinecke2002,Roepke2005}, which uses a level-set based prescription of
the flame front \citep{Reinecke1999} together with PPM hydrodynamics
\citep{PPM1984} and a bilinear interpolation of the Timmes equation of state
\citep[EoS;][]{Timmes1999}. The gravitational field is described using a
Newtonian potential. Where Coulomb corrections are considered in the EoS it is
by the formulation of \citet{Potekhin2000}. The computational grid consists of
a uniform Cartesian grid that expands to track the evolution of the flame front
and is nested within a non-uniform Cartesian grid that is designed to follow
the expansion of the whole star \citep{Roepke2006}.

The flame speeds are taken from the fitting formulae of \citet{Timmes1992} in
the laminar regime. A subgrid model for turbulence by \citet{Schmidt2006} is
used to calculate the kinetic energy of the fluid on length scales smaller than
are resolvable given a finite numerical resolution. The turbulent velocities
from this subgrid model are then used to give the flame speed in the turbulent
regime.

The nuclear energy release is determined from tabulated NSE abundances, which
require an iterative method in order to solve simultaneously for the
temperature and the composition. The NSE abundances are tabulated with the
density of the fuel being burned as the independent variable. The tables are
produced by iteratively writing the abundance tables from post-processing
nuclear network calculations and simulating the deflagration in 2D (the same
method as used by \citealp{Fink2010} for detonations and \citealp{Ohlmann2014}
for deflagrations). Thermal neutrino losses are also included in the
simulations using the formulae by \citet{Itoh1996}, as described in
\citet{Seitenzahl2015}.

The veracity of the assumption that the burning proceeds as a deflagration, as
opposed to a detonation, is an interesting subject.  The pure detonation of
Chandrasekhar-mass CO WDs was ruled out as the predominant explosion scenario
for type Ia SNe because the intermediate-mass elements (IMEs) are underproduced
\citep{Arnett1971}. Neither have such explosions been observed: there is always
a signature of IMEs in the nebular spectra of type Ia SNe. However, since ECSNe
and the AIC of ONe WDs are expected to be much less frequent events and whose
observational characteristics are not well known, that their burning mode is a
detonation cannot be completely ruled out using the same arguments.

\subsection{Initial flame surface structure}
\label{sec:ignition-geometry}

The initial flame structure is a critical input of the 3D hydrodynamic
simulations. 1D stellar models of the pre-deflagration evolution of ECSN
progenitor stars and accreting ONe WDs suggest that the deflagration is ignited
at a single point in the centre of the star
\citep{Nomoto1984,Nomoto1987,Schwab2015}. It is of course not possible to
resolve a point in any geometrical sense, and instead the initial flame shape
should be something that is representative of the deflagration once it has
moved outwards from the centre by a small distance. From a numerical
standpoint, it must have moved out far enough that it is able to be resolved by
the resolution of the computational grid. This leads to a degree of speculation
with regards to how the flame surface has evolved during the time between its
inception and the time at which one can resolve it.

In the simulations presented in this work, the flame surface at the beginning
of the simulation is constructed as 300 spherical bubbles uniformly distributed
within a sphere of radius 50~km with its origin at the centre of the ONe core.
This represents the surface of a single flame perturbed with small-scale modes.
For the model with initial central density
$\rho_\mathrm{c}^\mathrm{ini}=10^{9.9}$~g~cm$^{-3}$, there is
$4.8\times10^{-6}~M_\odot$ of fuel inside the initial flame, that is
transformed into ash during the first time step.  The outcome of these
simulations is expected to be sensitive to the initial geometry of the flame
\citep[see, e.g.,][]{Seitenzahl2013}, however a centrally confined
ignition with small, but resolvable, perturbations is a sensible first approach
(see the Summary and Conclusions of this work for a brief discussion).

\subsection{Deleptonisation of the NSE ashes}
\label{sec:yedot}

\begin{table*}
\setlength{\tabcolsep}{12pt}
\centering

\caption{Summary of the 3D O deflagration simulations. The horizontal dashed
line separates those models above it, in which enough material becomes unbound
from the ONe core/WD that the remaining bound remnant has a mass below the
effective Chandrasekhar limit $M_\mathrm{Ch}$, from those below (i.e. the H
series of models) which will collapse to form a neutron star.  The quoted Fe
masses are actually the sum of all the Fe-group elements.}

\begin{tabular}{l c c c c@{\hskip 6pt}c c@{\hskip 6pt}c c c c}
\toprule

id. & res. & $^a\log_{10}\rho_\mathrm{c}^\mathrm{ini}$ & $^b$CC &
$^cM_\mathrm{rem}$ & $^dM_\mathrm{rem}^\mathrm{Fe}$ & $^eM_\mathrm{ej}$ &
$^fM_\mathrm{ej}^\mathrm{Fe}$ & $^g\langle Y_\mathrm{e,rem}\rangle$ &
$^hM_\mathrm{Ch}^\mathrm{eff}$ & $^i\Delta x$ \\

 & & (g~cm$^{-3})$ & (Y/N) & \multicolumn{2}{c}{($M_\odot$)} &
\multicolumn{2}{c}{($M_\odot$)} & & ($M_\odot$) & (km)\\

\midrule

G13    & $256^3$ & 9.90 & N & 0.647 & 0.173 & 0.741 & 0.231 & 0.491 & 1.384 &  0.870 \\
G14    & $512^3$ & 9.90 & N & 0.438 & 0.115 & 0.951 & 0.362 & 0.491 & 1.381 &  0.427 \\
G15    & $256^3$ & 9.90 & Y & 1.212 & 0.223 & 0.177 & 0.047 & 0.493 & 1.392 &  0.870 \\
J01    & $256^3$ & 9.95 & N & 0.631 & 0.171 & 0.768 & 0.233 & 0.491 & 1.379 &  0.870 \\
J02    & $256^3$ & 9.95 & Y & 1.291 & 0.226 & 0.104 & 0.025 & 0.493 & 1.392 &  0.870 \\
\hdashline
H01$^*$ & $256^3$ & 10.3 & N & 1.401 & 0.022 & 0.000 & 0.000 & 0.486 & 1.356 &  0.870 \\

\bottomrule
\end{tabular}

\raggedright
$^*${\tiny These diagnostics were calculated for H01
at $t\approx330$~ms, at which time the simulation was stopped because the
maximum density reached $\sim10^{11}$~g~cm$^{-3}$, exhibiting clear signs of
core collapse. Thus, $M_\mathrm{rem}^\mathrm{Fe}$ for the {\sc H01} model is
the mass of Fe-group elements in the core after $\sim330$~ms}.

\tiny
\begin{tabular}{ll}
$^a$~Ignition (central) density of ONe core at ignition of O deflagration. & 
$^b$~Coulomb corrections included in EoS \\
$^c$~Total mass of bound ONeFe WD remnant & $^d$~Mass of Fe-group elements in bound remnant \\
$^e$~Total ejected mass & $^f$~Mass of ejected Fe-group elements \\
$^g$~Average electron fraction of bound remnant & $^h$~Effective Chandrasekhar mass of bound remnant \\
$^i$~Initial cell size for inner (flame) mesh \\
\end{tabular}

\label{tab:model-props}
\end{table*}

It is not only electron capture that is important for the evolution of $Y_{\rm
e}$ in the ashes of the flame: $\beta$-decay and positron capture rates are
also significant under the thermodynamic conditions and timescales experienced
during the O deflagration. Thus, it was not possible to use the
state-of-the-art e$^-$-capture rates by \citet[][J10; as in
\citealp{Takahashi2013}]{Juodagalvis2010} that are used in core-collapse
supernova simulations. This is because (a) J10 do not provide $\beta$-decay
rates (which are not significant in core-collapse supernovae) and (b) using
inconsistent e$^-$-capture and $\beta$-decay rates (e.g. e$^-$-capture rates
from J10 and $\beta$-decay rates from \citealp{LMP2001}) would upset the
detailed balance and misrepresent the deleptonisation rate in the simulation.

Instead, the evolution of $Y_\mathrm{e}$ is followed using its time derivative
$\dot{Y}_\mathrm{e}(\rho,T,Y_\mathrm{e}$), which is interpolated from
pre-processed tables with $T$, $\rho$ and $Y_{\rm e}$ as independent variables.
The tables were constructed in a similar manner to \citet{Seitenzahl2009}: for
each point in the 3-dimensional parameter space of the independent variables
the NSE equations are solved \citep{Seitenzahl2009,Pakmor2012} and the
contribution of each nuclear species to the change in $Y_\mathrm{e}$ is
accounted for by folding its abundance with the relevant weak reaction rates.

The sources of the electron-capture reaction rates that were used are shown in
Fig.~\ref{fig:rate-sources}. The $\beta$-decay (inverse) rates were taken from
the same sources as their electron-capture counterparts, to ensure consistency.
For protons, neutrons and $pf$-shell nuclei the shell model rates of
\citet[][LMP]{LMP2001} were used where available. For the $sd$-shell, the rates
of \citet[][ODA]{ODA94} were used (also shell model calculations) where
available. Where LMP and ODA rates were not available, the rates from
\citet[][FFN]{FFNweak1985} were used. Where rates were still missing, the
choice falls back to the QRPA rates of \citet[][NKK]{Nabi2004}. All other rates
were computed in a similar manner to the approximations described in
\citet{Arcones2010} and \citet{Sullivan2016}. The neutrino luminosity
$\epsilon_\nu$ is also computed and tabulated in the same manner as
$\dot{Y}_\mathrm{e}$ and is used as a sink term in the energy equation. This is
a valid approximation until the density reaches approximately
$10^{11}$~g~cm$^{-3}$, where neutrino interactions with matter become
non-negligible. Such high densities would be realised only in the case of a
core collapse event.


\section{Results}
\label{sec:results}
\label{sec:bound-remnants}

\begin{figure*}
\centering
\includegraphics[width=1.\linewidth]{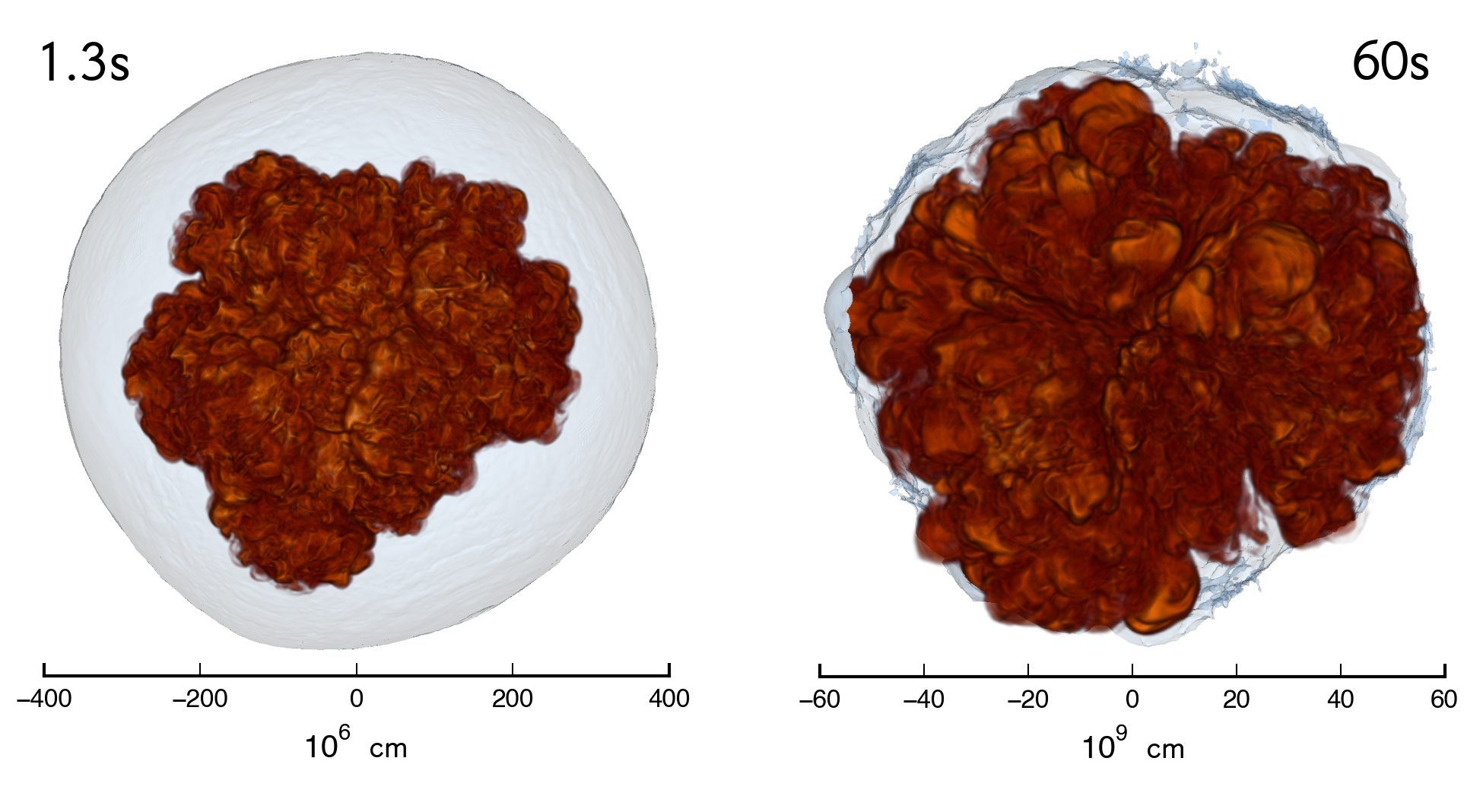}

\caption{Mass fraction of Fe-group elements in simulation G14 (see
Table~\ref{tab:model-props} for details) at 1.3s (left panel) and 60s (right
panel) of simulated time. The greyish-blue contour is the surface of the ONe
core, which shows a distinct aspherical deformation. Interior to the
$0.951~M_\odot$ of ejecta (which consists of $0.362~M_\odot$ of Fe-group
elements) remains a bound remnant of $0.438~M_\odot$ consisting of a mixture of
O, Ne and $0.115~M_\odot$ of Fe-group elements.}

\label{fig:G14nifs-60s}
\end{figure*}

A number of simulations were performed in either $256^3$ or $512^3$ resolution
for three initial ONe core structures with
$\log_{10}\rho_\mathrm{c}^\mathrm{ini}=9.9$ (G series), 9.95 (J series) and
10.3 (H series). Some of the simulations included the effect of Coulomb
corrections in the EoS. A summary of the key simulations is provided in
Table~\ref{tab:model-props} along with some diagnostic information.

\subsection{Outcomes of the simulations}

There is a clear distinction between the lower density G and J models and the H
models, which had the highest ignition densities. This is shown in
Fig.~\ref{fig:rhomax-yemin-t}, where the maximum density is plotted as a
function of time. In the first 5 seconds, the maximum density in G14 and J02
drops by several orders of magnitude owing to expansion caused by the release
of nuclear binding energy.  H01, with the highest initial density, on the other
hand, experiences only contraction.  Within the first few hundred milliseconds
the maximum density increases by a factor of 5, reaching almost
$10^{11}$~g~cm$^{-3}$. At these high densities, the interaction of neutrinos
with matter (which is not accounted for in these simulations) should become
significant and therefore the simulation could not be continued with the
methods and assumptions that were used. The result of the H01 simulation is,
however, rather indicative that a core-collapse event will be the fate of such
a star. On the other hand, the G and J models -- which are based on the current
understanding of the $^{20}$Ne electron-capture process
\citep{Martinez-Pinedo2014,Schwab2015} -- are a kind of thermonuclear explosion
in which a bound ONeFe compact remnant is produced. Although the models by
\citet{Schwab2015} include the most accurate treatment of the electron-capture
processes, an additional factor in determining the ignition density of the O
deflagration in ONe cores is the onset of semiconvection. The dispersion
relation for this kind of overstable oscillatory convection was derived by
\citet{Kato1966}, giving a growth rate of the overstable oscillations. A
simplification to the solution of Kato's equation by \citet[][but see also
\citealp{Shibahashi1976}]{Langer1983} forms the basis of the parameterised
diffusion approximation to semiconvective mixing found in many modern stellar
evolution codes. \citet{Schwab2015} showed with a timescale argument that
Langer's formulation of semiconvective mixing would likely not (i.e., within
the range of typically-used values of the free parameter $\alpha$) impact upon
the ignition density of the O deflagration.  \citet{Takahashi2013} on the other
hand found an ignition density of $\sim3\times10^{10}$~g~cm$^{-3}$ while
accounting for semiconvective mixing using the time-dependent mixing
formulation described in \citet{Unno1967}. This may have been an artefact of
the under-resolved tabulations of the $^{20}$Ne and $^{20}$F electron-capture
rates by \citet{ODA94} used in \citet{Takahashi2013}.  The importance of
semiconvection during the $^{20}$Ne electron-capture preceding the
thermonuclear runaway is an interesting problem in itself, however it is one
that is outside of the scope of the present work. The calculations of
\citet{Schwab2015} are taken to be the current standard until further work into
the reaction rates or semiconvective instability are undertaken. The
simulations performed in this work represent the two extreme cases of
semiconvection: in the case of inefficient semiconvection, the G or J series of
simulations are the most realistic (depending on the strength of the second
forbidden transition from $^{20}$Ne to $^{20}$F); in the case of semiconvection
being so efficient that fully developed convection is established over a short
time scale, the H series of simulations would probably be closer to reality.

Another critical quantity in these simulations is the minimum electron fraction
$Y_\mathrm{e}^\mathrm{min}$, which is also shown in
Fig.~\ref{fig:rhomax-yemin-t} (solid blue lines with circular glyphs). In the
G14 (J02) model $Y_\mathrm{e}^\mathrm{min}$ does not go below about 0.40
(0.41), and after a few tens of milliseconds begins to increase again as the
density drops and $\beta$-decays become more prevalent. As in the case of the
maximum density, \yemin~behaves differently for the H models than for the G and
J models: The decrease in pressure caused by the marked reduction in the number
of electrons induces a contraction of the ONe core/WD in the H01 simulation
which in turn increases the temperature and density, accelerating the rate of
deleptonisation in a runaway process. Another important consequence of the
reduction in electron fraction and the increase in temperature during the
associated contraction is the adjustment of the NSE state. At
$Y_\mathrm{e}=0.3$, $T=15$~GK and $\log_{10}\rho=10.5$ the equilibrium state
consists of free neutrons, $\alpha$ particles, intermediate-mass elements and
Fe-group elements in roughly equal parts (Fig.~\ref{fig:nse-abunds}). The
internal energy of such a state is in fact lower than the internal energy of
the initial model with O and Ne at $T=5\times10^5$~K. This, too, results in a
gas pressure deficit and favours a core collapse event.

In about the first 150~ms of the H01 simulation, \yemin~becomes as low as 0.25
-- which was the lower bound of the \yedot~and $\epsilon_\nu$ tables that were
used -- and does not decrease any further. Even with the deleptonisation
curtailed because of the inadequate domain of the \yedot~tables that had been
computed as described in Section~\ref{sec:yedot}, with $Y_\mathrm{e}$ at the
upper limit of 0.25 the model still shows a clear indication of core collapse.

\subsection{The bound remnants: ONeFe white dwarfs}

A key result of the G and J simulations is that material becomes unbound from
the ONe core as a result of the release of nuclear energy during the
deflagration, leaving a bound remnant consisting of O and Ne ($\sim60-80\%$),
iron-group elements ($\sim20-40\%$), and intermediate-mass elements ($1-3\%$).
This is in rather good agreement with the 1D calculations of \citet{Isern1991},
even though there are some appreciable differences between the approaches of
this work and theirs. To summarize: this work uses a more sophisticated EoS,
laminar flame speeds from microscopic flame calculations \citep{Timmes1992},
modern weak reaction rates, a subgrid model of turbulence and is performed in
$4\pi$ geometry with 3D hydrodynamics. The mass fractions of O and Ne (0.65 and
0.35, respectively) used in the present work are more representative of the
actual composition of the ONe cores, compared to the abundances from
\citet{Miyaji1980} that were used by Isern et al.. In addition, from the
important work of \citet{Martinez-Pinedo2014} and \citet{Schwab2015}, the
present work benefits from tighter constraints on the ignition density of the
deflagration although the value of the ignition density itself has changed
little since the work of \citet{Isern1991} and \citet{Canal1992}. Despite the
significant differences between the present work and that of \citet{Isern1991}
and \citet{Canal1992}, our qualitative results are in rather good agreement:
Isern et al. also found bound WD remnants consisting of a mixture of O, Ne and
Fe-group elements. This is really quite a remarkable result.

\begin{figure}
\includegraphics[width=\linewidth, clip=true, trim = 
0mm 17mm 0mm 5mm]{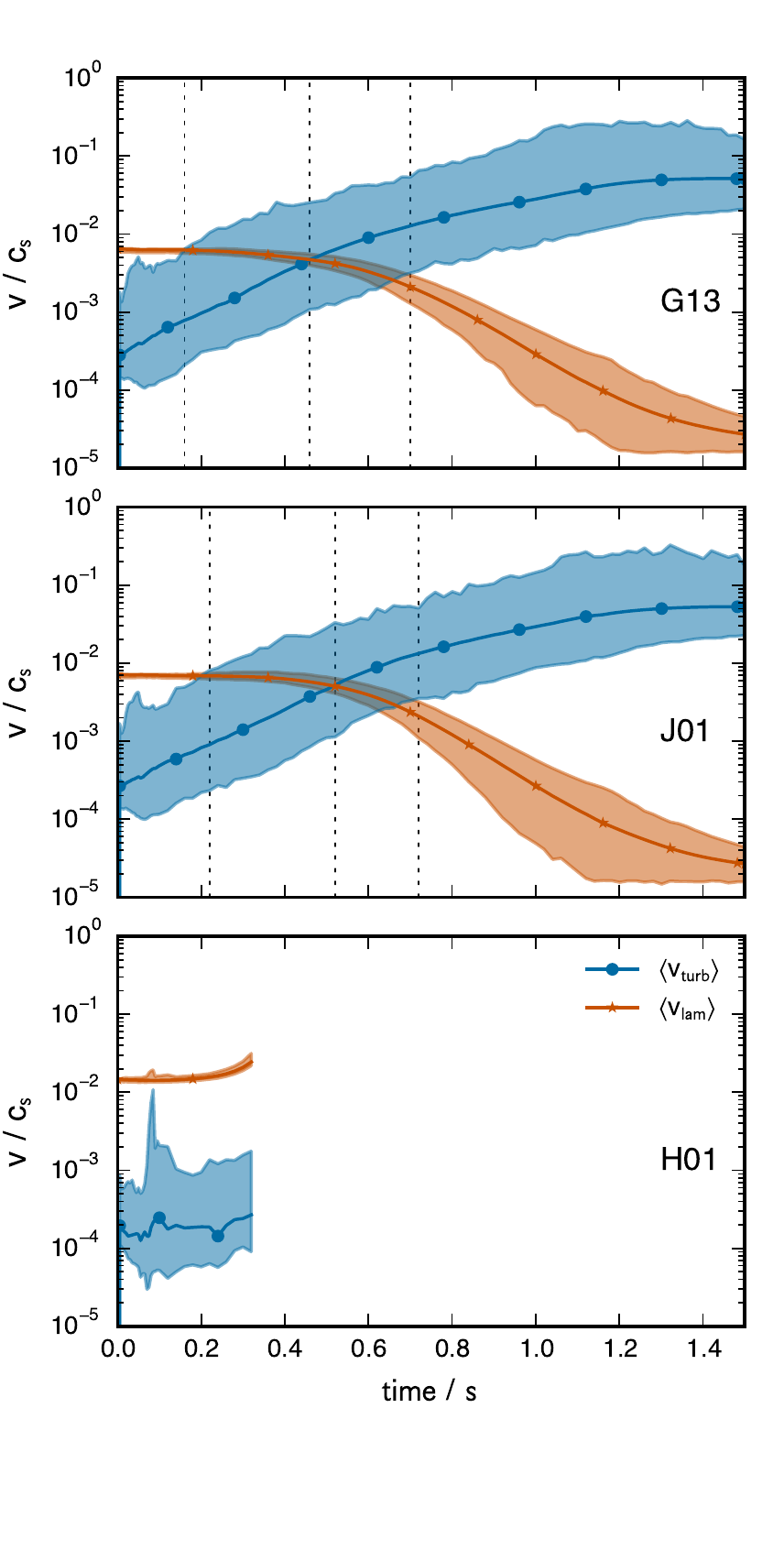} \\

\caption{Laminar and turbulent flame speeds as fractions of the sound speed
$c_s$ (i.e., Mach numbers) during the first 1.5 seconds of the {\sc G13}, {\sc
J01} and {\sc H01} simulations (see Table~\ref{tab:model-props}). The solid
lines show the average (mean) value over the entire burning front. The shaded
region extends from the minimum to the maximum speed. The vertical dashed lines
demarcate the times at which (a) the flame becomes turbulent for the first
time, (b) the mean turbulent flame speed exceeds the mean laminar flame speed
and (c) the flame is completely turbulent. In simulation {\sc J01}
($\log\rho_\mathrm{c}^\mathrm{ini}=9.95$), the flame becomes turbulent slightly
later than in {\sc G15} ($\log\rho_\mathrm{c}^\mathrm{ini}=9.90$), whereas in
{\sc H01} ($\log\rho_\mathrm{c}^\mathrm{ini}=10.3$) the flame never becomes
turbulent and the simulation shows a clear sign that it will collapse into a
neutron star. } 

\label{fig:flame-speeds1}
\end{figure}

\begin{figure}
\includegraphics[width=1.0\linewidth, clip=true, trim = 
0mm 7mm 0mm 0mm]{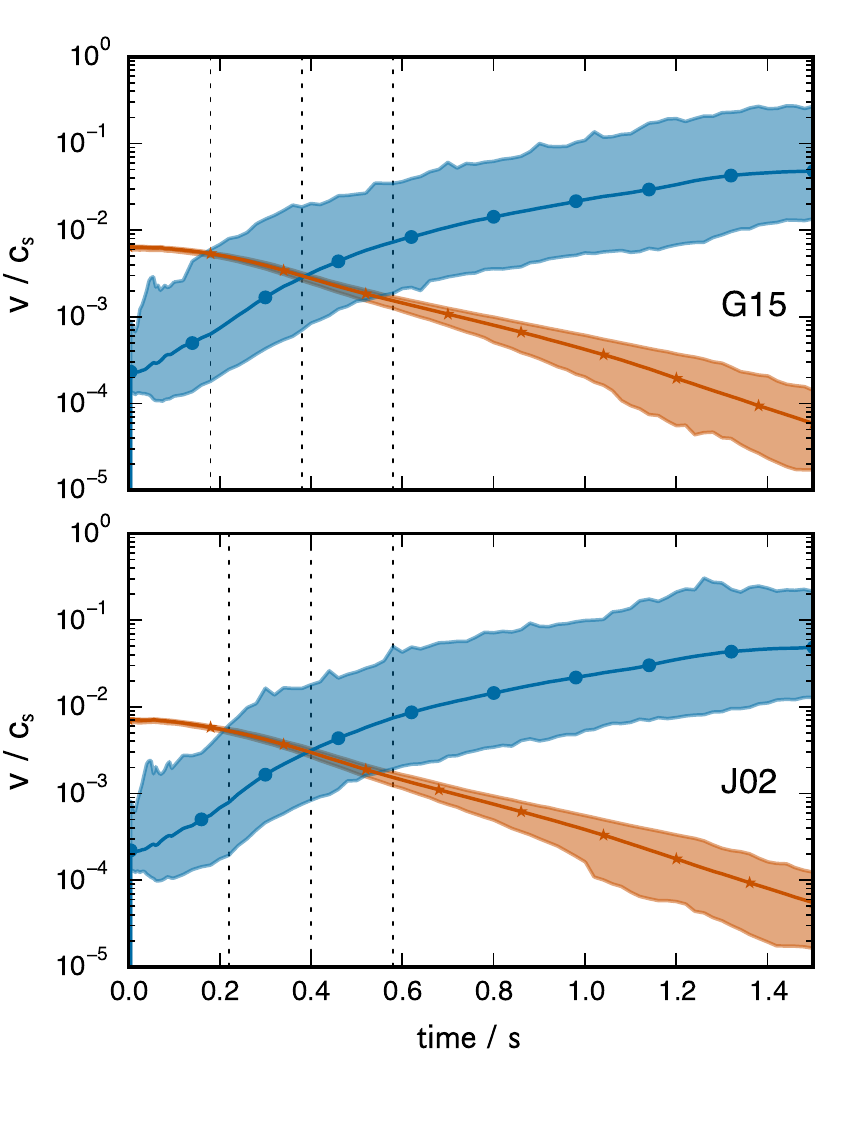} \\

\caption{Same as Fig.~\ref{fig:flame-speeds1} but for models {\sc G15} and
{\sc J02} ($\log\rho_\mathrm{c}^\mathrm{ini}=9.90$ and $9.95$, respectively),
in which Coulomb corrections are included in the EoS.}

\label{fig:flame-speeds2}
\end{figure}

The mass of the bound remnant is very sensitive to whether or not Coulomb
corrections are included in the EoS (see Table~\ref{tab:model-props}).
Simulations with and without the Coulomb corrections were performed.  Including
the corrections results in roughly a factor of 2 increase in the mass of the
bound remnant and a factor of $\sim5$ decrease in the ejecta mass. These are
significant changes and the sensitivity of the quantitative result to the
long-range coupling of the ideal components of the plasma being so great
motivates further scrutiny of the accuracy with which such Coulomb corrections
are treated. The laminar and turbulent flame speeds are shown as a function of
time for the $256^3$ simulations without and with Coulomb corrections to the
EoS in Figs.~\ref{fig:flame-speeds1} and \ref{fig:flame-speeds2}, respectively.
One can see that in the simulations where the internal energy -- and, hence,
the pressure -- of the plasma are reduced due to the Coulomb corrections the
flame takes longer to become dominated by turbulence. Quite why this seemingly
slight shift in the timing of the onset of the turbulent flame results in such
a marked decrease in the amount of mass that can reach escape velocity is a
detailed problem.

The G14 simulation ejected $0.951~M_\odot$ of material, of which
$0.362~M_\odot$ is composed of Fe-group elements, and leaves a bound remnant of
$0.438~M_\odot$. The bound remnant consists of $0.115~M_\odot$ of Fe-group
elements with the remaining mass comprised of a mixture of O and Ne in the same
65\%/35\% proportions of the initial composition of the ONe core/WD. Comparing
the numerical values in Table~\ref{tab:model-props} for models G13 ($256^3$)
and G14 ($512^3$) it is clear that the quantitative answer is not converged on
grid refinement.  However, there is a clear indication that the result is
qualitatively converged. In particular, the simulation at both resolutions
results in the partial ejection of the core material, leaving a bound compact
remnant.  Doubling the number of grid cells in each spatial dimension from 512
to 1024 and hence increasing the computational expense by a factor of 16 ($2^3$
more grid cells and a factor of 2 decrease in the Courant number) seems an
unnecessary expense at this time.

The J series of models with initial central densities
$\log_{10}\rho_\mathrm{c}^\mathrm{ini}=9.95$ display similar characteristics
and a similar trend to the G series of models with slightly lower ignition
density. All of the simulations eject a fraction of the core material and leave
a bound remnant with a mass not exceeding the effective Chandrasekhar mass
$M_\mathrm{Ch}^\mathrm{eff}$. The inclusion of Coulomb corrections to the EoS
approximately doubles the mass of the bound remnant, as in the G series.
Simulations without Coulomb corrections were not performed in $512^3$
resolution for the J series however the trend is expected to be the same as in
the G series; i.e. the mass of the bound remnant would be lower than in the
$256^3$ simulation.

\subsection{A multidimensional problem}

A volume rendering of the abundance of Fe-group elements at 60s of simulated
time in the G14 simulation (see Table~\ref{tab:model-props}) is shown in
Fig.~\ref{fig:G14nifs-60s}. The asymmetry of the deflagration and of the ejecta
is striking. This is a result of the growth of the R--T instability. Whether or
not the R--T instability is significant in O deflagrations was considered by
\citet{Timmes1992}, who have compared the minimum wavelength instability to the
thickness of the density inversion produced by the burning front before it is
erased by electron capture (their Fig.~9b). In their case A -- which has a
ratio of O/Ne that is closest to the core composition in the most recent models
of super-AGB stars using the $^{12}{\rm C}(\alpha,\gamma)^{16}{\rm O}$ rate of
\citet{Kunz2002} -- the instability grows for $\rho\lesssim1.1\times10^{10}$.
This is confirmed by the G14 simulation (and in fact by the entire G and J
series of simulations) of the present work, in which the density never exceeds
this value (Fig.~\ref{fig:rhomax-yemin-t}) and indeed R--T plumes can be seen
in the Fe-group abundances (Fig.~\ref{fig:G14nifs-60s}). Timmes \& Woosley also
predicted that for $\rho\gtrsim1.1\times10^{10}$, the R--T instability would be
suppressed and the core would collapse to form a neutron star.  This result is
confirmed by the H01 simulation of the present work (see
Table~\ref{tab:model-props} and Fig.~\ref{fig:rhomax-yemin-t}), which results
in core collapse. Indeed, the suppression of the R--T instability is observed
in the H01 simulation and the flame stays in the laminar regime.  The laminar
and turbulent flame speeds (see Section~\ref{sec:methods}) are shown in
Figs.~\ref{fig:flame-speeds1} and \ref{fig:flame-speeds2} for the $256^3$
simulations without and with EoS Coulomb corrections, respectively. The red and
blue lines are the mean values of the laminar and turbulent burning speeds,
respectively, as Mach numbers over the whole flame surface as a function of
time. The shaded regions show the total range of flame speeds over the surface
as a function of time. The vertical dashed lines demarcate the times at which
(from left to right) the flame becomes turbulent \emph{somewhere}, the mean
turbulent flame speed is higher than the mean laminar flame speed, and the
flame is turbulent \emph{everywhere}. These times are delayed slightly for
higher ignition density (compare, e.g., G13 and J01 or G15 and J02).
Interestingly, when Coulomb corrections are included in the EoS the flame takes
slightly longer to become turbulent anywhere on its surface, but less time is
needed for its speed to become completely dominated by turbulence (compare G13
and G15 or J01 and J02).

\begin{figure*}
\centering
\includegraphics[clip=true,trim=0mm 0mm 0mm 0mm, width=1.\linewidth]{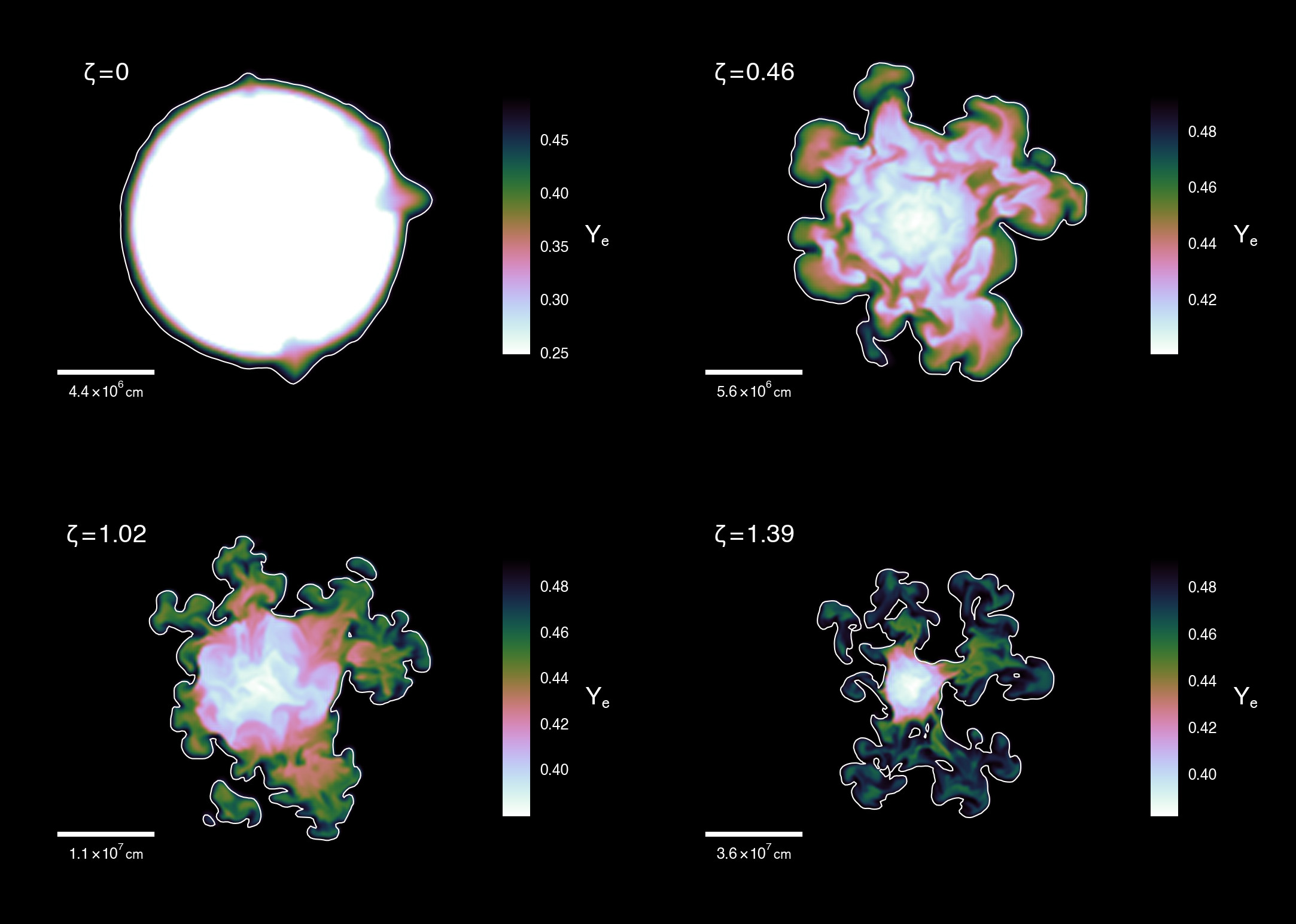}

\caption{Geometry of the flame surface for different values of the
asymmetry diagnostic parameter $\zeta$ (\ref{eq:zeta}). The plot for $\zeta=0$
is a snapshot of the H01 simulation at 320~ms of simulated time; the flame
surface is essentially a sphere due to the suppression of the Rayleigh-Taylor
instability. All of the other plots are of the J01 simulation. The time
evolution of $\zeta$ is shown in Fig.~\ref{fig:asymmetries1}; the asphericity
becomes more pronounced with time in the G and J series simulations. Each panel
also has a pseudo-colour plot of the electron fraction $Y_\mathrm{e}$.}

\label{fig:flame_vs_zeta}
\end{figure*}

\begin{figure}
\includegraphics[clip=true,trim=0mm 8.5mm 0mm 0mm, width=\linewidth]{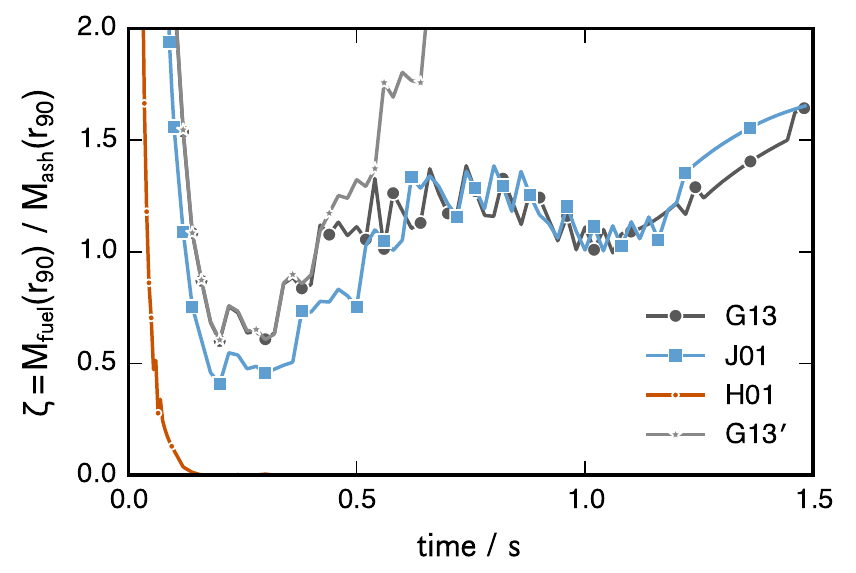} \\
\includegraphics[width=\linewidth]{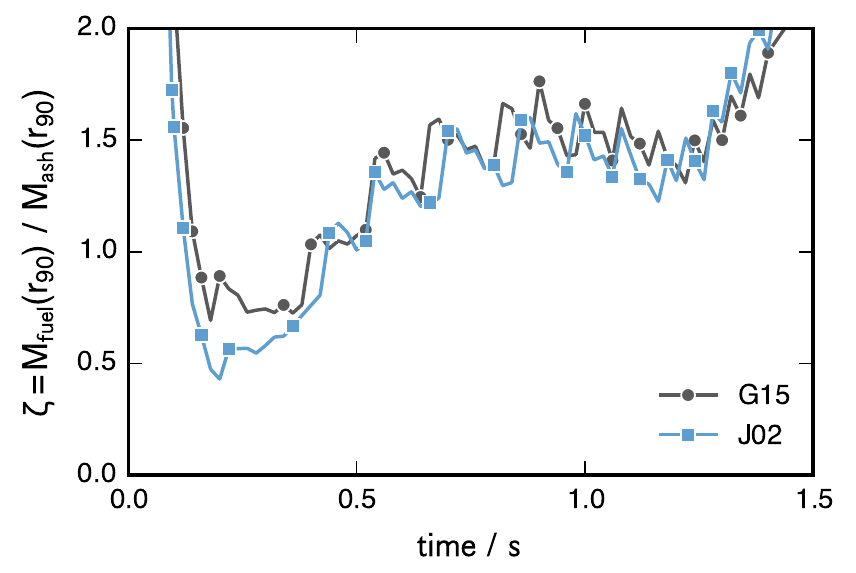}

\caption{Time evolution of the flame asymmetry diagnostic parameter
$\zeta$ for $256^3$ simulations from the G, J and H simulation series (see
Table~\ref{tab:model-props}). The bottom panel shows the simulations with EoS
Coulomb corrections, and the top panel shows the simulations without. The
asymmetry develops in much the same way regardless of the inclusion of Coulomb
corrections in the EoS, and is even slightly more pronounced when they are
included. Clearly, the Rayleigh-Taylor instability is suppressed in the highest
density simulation (H01) and the flame surface becomes essentially a sphere
($\zeta=0$). Fig.~\ref{fig:flame_vs_zeta} shows the severity of the asymmetry
for different values of $\zeta$, which is difficult to imagine a priori.}

\label{fig:asymmetries1}
\end{figure}

Understanding the impact of the R--T instability on the quantitative results
and qualitative outcome of the O deflagration in a differential sense is a
difficult -- if not impossible -- task, because the instability and its growth
are the direct result of solving the Euler equations, which is at the very
heart of multidimensional hydrodynamic simulations. However, we have defined an
instructive flame asymmetry diagnostic parameter
\begin{equation}
\zeta = \frac{M_\mathrm{fuel}(r_{90})}{M_\mathrm{ash}(r_{90})}\;,
\label{eq:zeta}
\end{equation}
which proves to be a rather robust metric that facilitates a quantitative
comparison of the degree of asymmetry of the flame surface resulting from the
growth or suppression of the R-T instability due to the relative buoyancy of
the ash to the fuel. In (\ref{eq:zeta}), $r_{90}$ is the radius of the sphere
containing 90\% of the ashes in the simulation by mass\footnote{The choice to
use 90\% is rather arbitrary, however the results vary very little indeed if
any value in the range 80--99\% is used.}.  $M_\mathrm{fuel}(r_{90})$ is then
the mass of fuel (material unburned by the deflagration) inside the radius
$r_{90}$ and similarly $M_\mathrm{ash}$ is the mass of ash inside that radius.
A value of $\zeta=0$ means that the flame surface is essentially a perfect
sphere, as one would find in a 1D representation (e.g.,
\citealp{Nomoto1991,Isern1991,Canal1992}). The degree of asymmetry
corresponding to the magnitude of $\zeta$ when $\zeta>0$ (i.e., anything other
than a sphere) is difficult to imagine and so in Fig.~\ref{fig:flame_vs_zeta}
plots of the flame surface for values of $\zeta=0$, 0.46, 1.02 and 1.39 are
shown, that are realised in the simulations. Since $\zeta$ becomes zero only in
the simulations with highest ignition density (H series), the corresponding
flame surface in Fig.~\ref{fig:flame_vs_zeta} (top left panel) is taken from
the H01 simulation.  All other panels of Fig.~\ref{fig:flame_vs_zeta} are from
the J01 simulation, for consistency.  At a value of $\zeta=0.46$ (top right
panel) the flame surface already shows significant deviations from spherical
symmetry. Therefore, Fig.~\ref{fig:flame_vs_zeta} serves as useful evidence
that for $\zeta\gtrsim0.46$ a spherically symmetric flame geometry would be a
rather poor approximation.  Even a value of $\zeta=0.1$ (not shown here) to the
eye looks distinctly non-spherical.  To put this into context, a comparison of
the flame asymmetry diagnostic parameter $\zeta$ for the three different
ignition densities as a function of time during the first 1.5~s (at which time
nuclear burning has ceased) is shown in Fig.~\ref{fig:asymmetries1}.  The
bottom panel shows $256^3$ simulations that include EoS Coulomb corrections and
the top panel those without. In both cases the G and J series of simulations
exhibit gross deviations from spherical symmetry, while in the high density H01
simulation the flame surface quickly becomes a sphere and remains as such.

\begin{figure*}
\centering
\includegraphics[clip=true,trim=25mm 0mm 0mm 0mm,width=.49\linewidth]{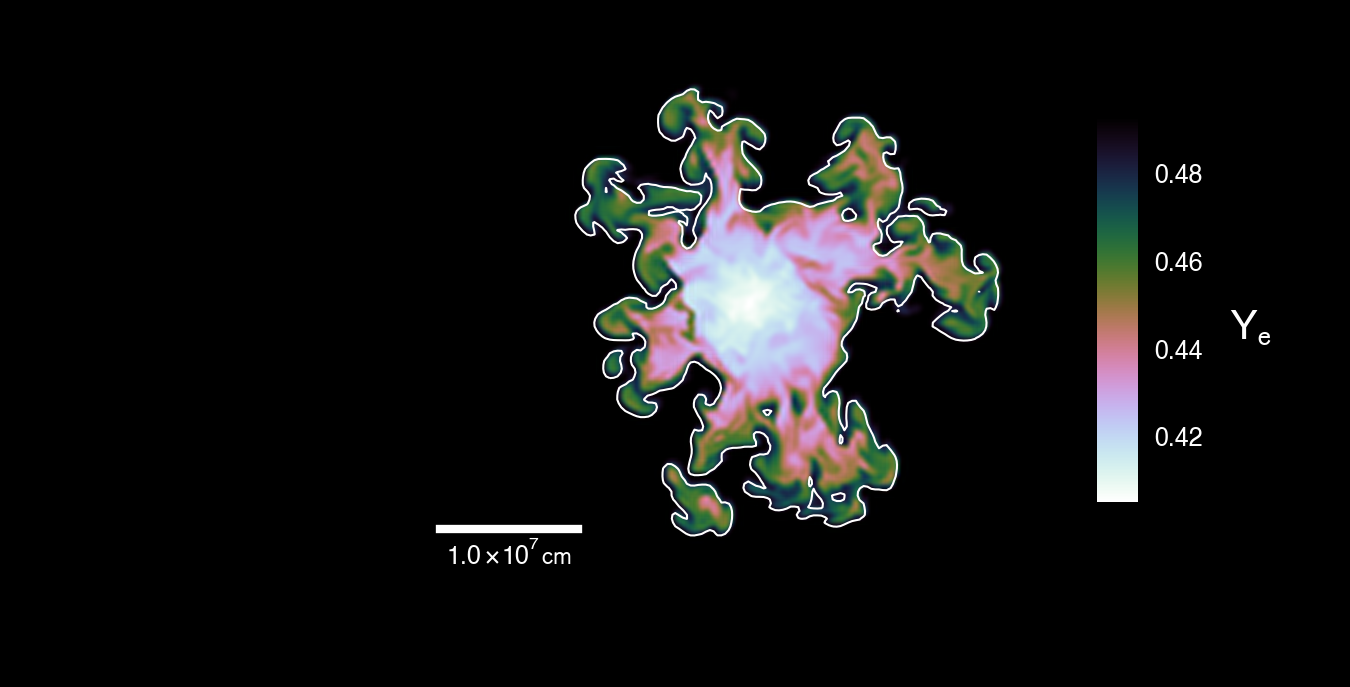}
\includegraphics[clip=true,trim=25mm 0mm 0mm 0mm,width=.49\linewidth]{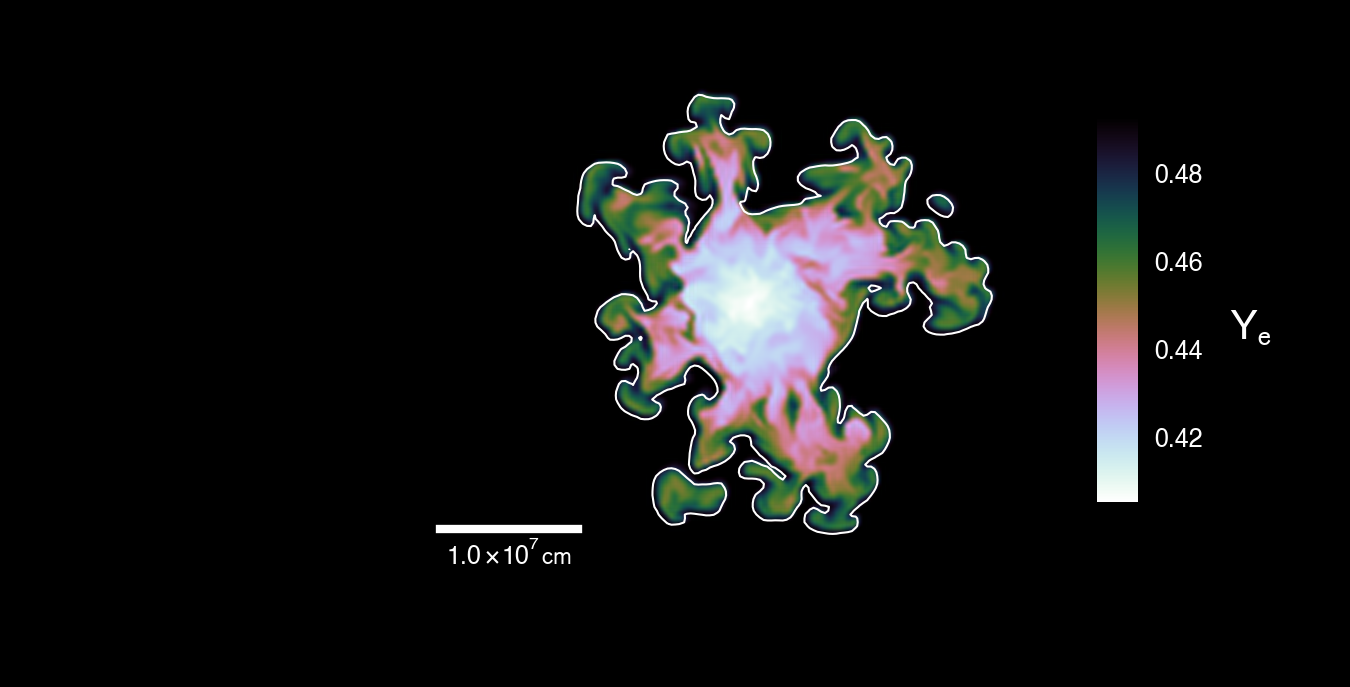}

\caption{Pseudo-color plot of the electron fraction, $Y_\mathrm{e}$, in a
slice through the x-y plane in two simulations at 500~ms of simulated time. The
right panel is the simulation G13 from Table~\ref{tab:model-props} and the left
panel is the same simulation with the turbulent burning speed set to zero, so
that the flame may only propagate conductively. The Rayleigh-Taylor fingers are
clearly visible in both cases. This shows that large-scale deviations from
spherical symmetry during the deflagration are not the result of turbulent
burning. The simulations look very similar indeed at this time: the mean
turbulent flame speed does not exceed the mean laminar flame speed in G13 until
460~ms (see Fig.~\ref{fig:flame-speeds1}). }

\label{fig:ye-slices}
\end{figure*}

If the flame had not become turbulent in the G and J series of models, and
instead had remained laminar (which is certainly only a hypothetical case), the
large-wavelength modes of the R--T instability that can be captured in 3D would
\emph{still} develop. This is shown in Fig.~\ref{fig:asymmetries1}, where the
line labelled G13$^\prime$ is the parameter $\zeta$ for a simulation that is
the same as G13 but with the turbulent flame speed set to zero, so that the
flame may only propagate via conduction. It behaves almost identically to G13
until around 0.5~s in that the flame surface becomes more asymmetric over time.
This is to be expected since the flame remains on average in the laminar regime
until about this time (as is shown in Fig.~\ref{fig:flame-speeds1}, top panel).
In Fig.~\ref{fig:ye-slices} the electron fraction and flame surface in the G13
simulation with only laminar flame speeds included (left panel) and with both
laminar and turbulent flame speeds included (right panel) are plotted at
500~ms. The snapshots do indeed look rather similar. The reason the asymmetry
parameter $\zeta$ for G13$^\prime$ rapidly increases after about 0.5~s is that
the material has expanded to densities below which the conductive deflagration
speeds have been calculated by \citet{Timmes1992}. The flame speed decreases
(as the red line in the top panel of Fig.~\ref{fig:flame-speeds1}) and, hence,
so does the rate of ash production. The ashes are still moved out to larger
radii by advection and expansion, however, resulting in a rapid increase in
$\zeta$.

Capturing the asphericity of the flame and the ashes using multidimensional
simulation techniques appears to be an integral part of studying
electron-capture supernovae. It may be possible to reproduce some, but
certainly not all, of the global characteristics of the simulations in this
work using one-dimensional methods with appropriately parameterised spherical
flame models. However, this would have to be proven and performing such 1D
simulations is beyond the scope of this work. In any case, the best-suited
parameters could almost certainly not be known a priori.


\section{Summary and Conclusions}
\label{sec:summary}

3D simulations of the O deflagration phase of ECSN progenitors and the AIC of
ONe WDs have been performed in $4\pi$ geometry. For ignition densities that we
consider to be representative in the light of recent updates to the $^{20}$Ne
electron-capture rate \citep{Martinez-Pinedo2014,Schwab2015}, collapse into a
neutron star does not occur. Instead, a thermonuclear explosion ejects a
portion of the degenerate core/WD, leaving behind a bound remnant consisting of
O, Ne and Fe-group elements, confirming the 1D simulations of
\citet{Isern1991}. What such an event would look like is certainly an
interesting question, and one that future work should attempt to address.  The
recent simulations by \citet{Schwab2015} are based on up-to-date nuclear
physics input. We therefore consider our low and intermediate ignition density
simulations as reference cases. Note, however, that \citet{Schwab2015} report
these to be lower limits. This is because of both the uncertain contribution of
the second forbidden transition to the electron-capture rate of $^{20}$Ne (as
discussed by Schwab et al.) and the uncertain impact of semi-convection.

Since the outcome of these deflagrations are known to be very sensitive to the
ignition density \citep[e.g.][]{Nomoto1991,Isern1991,Canal1992}, simulations
with rather high ignition density ($2\times10^{10}$~g~cm$^{-3}$) were also
computed.  These simulations show clear signs of core collapse, reaching a
maximum density of $10^{11}$~g~cm$^{-3}$ within 400~ms of the deflagration
being ignited.  Such a high density ignition may only be realised if
significant energy transport by convective motions takes place during the
evolution immediately preceding the thermonuclear runaway
\citep{Mochkovitch1984,Miyaji1987,Gutierrez1996}, although convection is likely
suppressed because of the stabilising gradient of mean molecular weight
produced during the electron-capture process \citep{Schwab2015}. In the case
that semiconvection during the $^{20}$Ne electron-capture phase is efficient
enough to destroy the gradient in mean molecular weight produced by the
electron captures the result will be fully-developed convection.  This is
likely not the case, however until a more rigorous examination is performed,
the contribution of semiconvective mixing to the ignition density remains
unclear.  It was not possible to follow the collapse of the core any further in
the highest density (H-series) models that reach such high densities owing to
the limitations of the EoS that was used and the omission of neutrino
interactions with matter from the present work.

It is also of course possible that updates to the microphysics in the G and J
series models could result in core collapse.  Indeed, the uncertainties in the
results of the simulations presented in this work are of course a product of
the uncertainties in the choice of input physics assumptions and those
introduced by the numerical implementation. An example worth highlighting is
the ignition geometry (see Section~\ref{sec:ignition-geometry}). In the present
work a centrally confined ignition was assumed, with small perturbations from
which the Rayleigh--Taylor instability can grow, should the time scales allow.
The perturbations were large enough to be able to be resolved on the
computational grid, so that any test of the sensitivity to numerical resolution
study makes sense. Even with these perturbations, cases where the R--T
instability could grow (G and J series) and cases where it was suppressed (H
series) were clearly distinguishable. It is quite difficult to imagine that the
flame would propagate outwards from a single central ignition point as a
perfect sphere, but that does not mean the magnitude of the perturbations used
in this work are realistic either.  The effect of varying the ignition geometry
in terms of number of ignition sparks, their location, asymmetries in the
distribution, etc., has been studied in detail for thermonuclear deflagrations
in Chandrasekhar-mass WDs at lower central densities \citep[see,
e.g.,][]{Garcia-Senz2005, Schmidt2006, Roepke2006, Roepke2007, Townsley2007,
Zingale2007, Jordan2008, Seitenzahl2011, Fink2014, Malone2014}.  It was shown
that they have generally a significant impact on the result in terms of nuclear
energy release and mass of the unbound material. Here, we have focused on
exploring the effect of the central density at ignition, keeping the geometry
of the initial flame fixed to a setup that favors collapse by restricting the
burning to the high-density central part of the star as long as possible. A
detailed investigation of the impact of the ignition geometry on the results
will be presented in a forthcoming publication.

There is also some uncertainty to be expected from the approximate treatment of
the flame using the level-set based approach, in which the flame structure
cannot be resolved.  The difference between resolving and not resolving the
flame structure is likely a rather small effect, but one should keep in mind
the marginality of this problem when considering how critical even small
uncertainties in the input physics are. The omission of general relativistic
corrections in the present work may introduce another small uncertainty that
could be important for the marginal phenomena
studied here.

The asphericity of the flame front in all but the highest density, collapsing
simulations, even if the flame were to remain in the laminar regime (which it
does not), is undeniable and 1D codes would be hard-pressed to be able to
reproduce the global properties of multidimensional hydrodynamic simulations.
This is especially true if the multidimensional simulations are not performed
first and therefore unable to inform the parameter choices for the 1D
simulations. This of course does not mean that 1D models of the O deflagration
are not useful, but it does mean that their predictive power is somewhat
restricted and depends upon the success of translating 3D simulation
diagnostics into 1D approximations.

These first simulations of the O deflagration in the progenitor stars of ECSNe
and the AIC of ONe WDs are a promising step towards understanding the nature of
these phenomena. The results present a strong motivation to constrain the
uncertainties in the modelling assumptions and test their impact on the outcome
of the ignition of O-burning in dense ONe cores/WDs. Ultimately, the
observational signatures of the various outcomes should be predicted and
compared with observations of both astronomical transients and supernova
remnants.

\begin{acknowledgements}

SJ is a fellow of the Alexander von Humboldt Foundation. SJ, FKR, RP, STO and
PVFE gratefully acknowledge support from the Klaus Tschira Foundation. RP
acknowledges support by the European Research Council under ERC-StG grant
EXAGAL 308037. IRS was funded by the Australian Research Council Laureate Grant
FL0992131. STO acknowledges support from Studienstiftung des deutschen Volkes.
SJ would like to personally thank Chris Fryer, Ken'ichi Nomoto, Nobuya
Nishimura, Gabriel Mart\'{i}nez-Pinedo, Raphael Hirschi and Wolfgang
Hillebrandt for helpful discussions and Frank Timmes for his comments on the
manuscript.

\end{acknowledgements}

%
%

\bibliographystyle{aa}
\bibliography{references}

\begin{thebibliography}{82}
\expandafter\ifx\csname natexlab\endcsname\relax\def\natexlab#1{#1}\fi

\bibitem[{{Arcones} {et~al.}(2010){Arcones}, {Mart{\'{\i}}nez-Pinedo},
  {Roberts}, \& {Woosley}}]{Arcones2010}
{Arcones}, A., {Mart{\'{\i}}nez-Pinedo}, G., {Roberts}, L.~F., \& {Woosley},
  S.~E. 2010, \aap, 522, A25

\bibitem[{{Arnett} {et~al.}(1971){Arnett}, {Truran}, \& {Woosley}}]{Arnett1971}
{Arnett}, W.~D., {Truran}, J.~W., \& {Woosley}, S.~E. 1971, \apj, 165, 87

\bibitem[{{Canal} {et~al.}(1992){Canal}, {Isern}, \& {Labay}}]{Canal1992}
{Canal}, R., {Isern}, J., \& {Labay}, J. 1992, \apjl, 398, L49

\bibitem[{{Colella} \& {Woodward}(1984)}]{PPM1984}
{Colella}, P. \& {Woodward}, P.~R. 1984, Journal of Computational Physics, 54,
  174

\bibitem[{{Denissenkov} {et~al.}(2013){Denissenkov}, {Herwig}, {Truran}, \&
  {Paxton}}]{Denissenkov2013cflame}
{Denissenkov}, P.~A., {Herwig}, F., {Truran}, J.~W., \& {Paxton}, B. 2013,
  \apj, 772, 37

\bibitem[{{Dominguez} {et~al.}(1993){Dominguez}, {Tornambe}, \&
  {Isern}}]{Dominguez1993}
{Dominguez}, I., {Tornambe}, A., \& {Isern}, J. 1993, \apj, 419, 268

\bibitem[{{Dunstall} {et~al.}(2015){Dunstall}, {Dufton}, {Sana}, {Evans},
  {Howarth}, {Sim{\'o}n-D{\'{\i}}az}, {de Mink}, {Langer}, {Ma{\'{\i}}z
  Apell{\'a}niz}, \& {Taylor}}]{Dunstall2015}
{Dunstall}, P.~R., {Dufton}, P.~L., {Sana}, H., {et~al.} 2015, \aap, 580, A93

\bibitem[{{Farmer} {et~al.}(2015){Farmer}, {Fields}, \& {Timmes}}]{Farmer2015}
{Farmer}, R., {Fields}, C.~E., \& {Timmes}, F.~X. 2015, \apj, 807, 184

\bibitem[{{Fink} {et~al.}(2014){Fink}, {Kromer}, {Seitenzahl},
  {Ciaraldi-Schoolmann}, {R{\"o}pke}, {Sim}, {Pakmor}, {Ruiter}, \&
  {Hillebrandt}}]{Fink2014}
{Fink}, M., {Kromer}, M., {Seitenzahl}, I.~R., {et~al.} 2014, \mnras, 438, 1762

\bibitem[{{Fink} {et~al.}(2010){Fink}, {R{\"o}pke}, {Hillebrandt},
  {Seitenzahl}, {Sim}, \& {Kromer}}]{Fink2010}
{Fink}, M., {R{\"o}pke}, F.~K., {Hillebrandt}, W., {et~al.} 2010, \aap, 514,
  A53

\bibitem[{{Fischer} {et~al.}(2010){Fischer}, {Whitehouse}, {Mezzacappa},
  {Thielemann}, \& {Liebend{\"o}rfer}}]{Fischer2010}
{Fischer}, T., {Whitehouse}, S.~C., {Mezzacappa}, A., {Thielemann}, F.-K., \&
  {Liebend{\"o}rfer}, M. 2010, \aap, 517, A80

\bibitem[{{Fuller} {et~al.}(1980){Fuller}, {Fowler}, \& {Newman}}]{FFN80}
{Fuller}, G.~M., {Fowler}, W.~A., \& {Newman}, M.~J. 1980, \apjs, 42, 447

\bibitem[{{Fuller} {et~al.}(1982){Fuller}, {Fowler}, \& {Newman}}]{FFN1982a}
{Fuller}, G.~M., {Fowler}, W.~A., \& {Newman}, M.~J. 1982, \apj, 252, 715

\bibitem[{{Fuller} {et~al.}(1985){Fuller}, {Fowler}, \& {Newman}}]{FFNweak1985}
{Fuller}, G.~M., {Fowler}, W.~A., \& {Newman}, M.~J. 1985, \apj, 293, 1

\bibitem[{{Garc{\'{\i}}a-Senz} \& {Bravo}(2005)}]{Garcia-Senz2005}
{Garc{\'{\i}}a-Senz}, D. \& {Bravo}, E. 2005, \aap, 430, 585

\bibitem[{{Gutierrez} {et~al.}(1996){Gutierrez}, {Garcia-Berro}, {Iben},
  {Isern}, {Labay}, \& {Canal}}]{Gutierrez1996}
{Gutierrez}, J., {Garcia-Berro}, E., {Iben}, Jr., I., {et~al.} 1996, \apj, 459,
  701

\bibitem[{{Hashimoto} {et~al.}(1993){Hashimoto}, {Iwamoto}, \&
  {Nomoto}}]{Hashimoto1993}
{Hashimoto}, M., {Iwamoto}, K., \& {Nomoto}, K. 1993, \apjl, 414, L105

\bibitem[{{Isern} {et~al.}(1991){Isern}, {Canal}, \& {Labay}}]{Isern1991}
{Isern}, J., {Canal}, R., \& {Labay}, J. 1991, \apjl, 372, L83

\bibitem[{{Itoh} {et~al.}(1996){Itoh}, {Hayashi}, {Nishikawa}, \&
  {Kohyama}}]{Itoh1996}
{Itoh}, N., {Hayashi}, H., {Nishikawa}, A., \& {Kohyama}, Y. 1996, \apjs, 102,
  411

\bibitem[{{Jones} {et~al.}(2014){Jones}, {Hirschi}, \& {Nomoto}}]{Jones2014}
{Jones}, S., {Hirschi}, R., \& {Nomoto}, K. 2014, \apj, 797, 83

\bibitem[{{Jones} {et~al.}(2013){Jones}, {Hirschi}, {Nomoto}, {Fischer},
  {Timmes}, {Herwig}, {Paxton}, {Toki}, {Suzuki}, {Mart{\'{\i}}nez-Pinedo},
  {Lam}, \& {Bertolli}}]{Jones2013}
{Jones}, S., {Hirschi}, R., {Nomoto}, K., {et~al.} 2013, \apj, 772, 150

\bibitem[{{Jones} {et~al.}(2016){Jones}, {Ritter}, {Herwig}, {Fryer},
  {Pignatari}, {Bertolli}, \& {Paxton}}]{Jones2016}
{Jones}, S., {Ritter}, C., {Herwig}, F., {et~al.} 2016, \mnras, 455, 3848

\bibitem[{{Jordan} {et~al.}(2008){Jordan}, {Fisher}, {Townsley}, {Calder},
  {Graziani}, {Asida}, {Lamb}, \& {Truran}}]{Jordan2008}
{Jordan}, IV, G.~C., {Fisher}, R.~T., {Townsley}, D.~M., {et~al.} 2008, \apj,
  681, 1448

\bibitem[{{Juodagalvis} {et~al.}(2010){Juodagalvis}, {Langanke}, {Hix},
  {Mart{\'{\i}}nez-Pinedo}, \& {Sampaio}}]{Juodagalvis2010}
{Juodagalvis}, A., {Langanke}, K., {Hix}, W.~R., {Mart{\'{\i}}nez-Pinedo}, G.,
  \& {Sampaio}, J.~M. 2010, Nuclear Physics A, 848, 454

\bibitem[{{Kato}(1966)}]{Kato1966}
{Kato}, S. 1966, \pasj, 18, 374

\bibitem[{{Kitaura} {et~al.}(2006){Kitaura}, {Janka}, \&
  {Hillebrandt}}]{Kitaura2006}
{Kitaura}, F.~S., {Janka}, H.-T., \& {Hillebrandt}, W. 2006, \aap, 450, 345

\bibitem[{{Knigge} {et~al.}(2011){Knigge}, {Coe}, \&
  {Podsiadlowski}}]{Knigge2011}
{Knigge}, C., {Coe}, M.~J., \& {Podsiadlowski}, P. 2011, \nat, 479, 372

\bibitem[{{Kunz} {et~al.}(2002){Kunz}, {Fey}, {Jaeger}, {Mayer}, {Hammer},
  {Staudt}, {Harissopulos}, \& {Paradellis}}]{Kunz2002}
{Kunz}, R., {Fey}, M., {Jaeger}, M., {et~al.} 2002, \apj, 567, 643

\bibitem[{{Lam} {et~al.}(2014){Lam}, {Mart{\'{\i}}nez-Pinedo}, {Langanke},
  {Jones}, {Hirschi}, {Zegers}, \& {Brown}}]{Lam2014}
{Lam}, Y.~H., {Mart{\'{\i}}nez-Pinedo}, G., {Langanke}, K., {et~al.} 2014, in
  European Physical Journal Web of Conferences, Vol.~66, European Physical
  Journal Web of Conferences, 7011

\bibitem[{{Langanke} \& {Mart{\'{\i}}nez-Pinedo}(2001)}]{LMP2001}
{Langanke}, K. \& {Mart{\'{\i}}nez-Pinedo}, G. 2001, Atomic Data and Nuclear
  Data Tables, 79, 1

\bibitem[{{Langer} {et~al.}(1983){Langer}, {Fricke}, \&
  {Sugimoto}}]{Langer1983}
{Langer}, N., {Fricke}, K.~J., \& {Sugimoto}, D. 1983, \aap, 126, 207

\bibitem[{{Lau} {et~al.}(2012){Lau}, {Gil-Pons}, {Doherty}, \&
  {Lattanzio}}]{Lau2012}
{Lau}, H.~H.~B., {Gil-Pons}, P., {Doherty}, C., \& {Lattanzio}, J. 2012, \aap,
  542, A1

\bibitem[{{Lecoanet} {et~al.}(2016){Lecoanet}, {Schwab}, {Quataert},
  {Bildsten}, {Timmes}, {Burns}, {Vasil}, {Oishi}, \& {Brown}}]{Lecoanet2016}
{Lecoanet}, D., {Schwab}, J., {Quataert}, E., {et~al.} 2016, ArXiv e-prints
  [\eprint[arXiv]{1603.08921}]

\bibitem[{{Malone} {et~al.}(2014){Malone}, {Nonaka}, {Woosley}, {Almgren},
  {Bell}, {Dong}, \& {Zingale}}]{Malone2014}
{Malone}, C.~M., {Nonaka}, A., {Woosley}, S.~E., {et~al.} 2014, \apj, 782, 11

\bibitem[{{Mart{\'{\i}}nez-Pinedo} {et~al.}(2014){Mart{\'{\i}}nez-Pinedo},
  {Lam}, {Langanke}, {Zegers}, \& {Sullivan}}]{Martinez-Pinedo2014}
{Mart{\'{\i}}nez-Pinedo}, G., {Lam}, Y.~H., {Langanke}, K., {Zegers}, R.~G.~T.,
  \& {Sullivan}, C. 2014, \prc, 89, 045806

\bibitem[{{Miyaji} \& {Nomoto}(1987)}]{Miyaji1987}
{Miyaji}, S. \& {Nomoto}, K. 1987, \apj, 318, 307

\bibitem[{{Miyaji} {et~al.}(1980){Miyaji}, {Nomoto}, {Yokoi}, \&
  {Sugimoto}}]{Miyaji1980}
{Miyaji}, S., {Nomoto}, K., {Yokoi}, K., \& {Sugimoto}, D. 1980, \pasj, 32, 303

\bibitem[{{Mochkovitch}(1984)}]{Mochkovitch1984}
{Mochkovitch}, R. 1984, in NATO Advanced Science Institutes (ASI) Series C,
  Vol. 134, NATO Advanced Science Institutes (ASI) Series C, ed. D.~{Bancel} \&
  M.~{Signore}, 125

\bibitem[{{Moriya} {et~al.}(2014){Moriya}, {Tominaga}, {Langer}, {Nomoto},
  {Blinnikov}, \& {Sorokina}}]{Moriya2014}
{Moriya}, T.~J., {Tominaga}, N., {Langer}, N., {et~al.} 2014, \aap, 569, A57

\bibitem[{{Nabi} \& {Klapdor-Kleingrothaus}(2004)}]{Nabi2004}
{Nabi}, J.-U. \& {Klapdor-Kleingrothaus}, H.~V. 2004, Atomic Data and Nuclear
  Data Tables, 88, 237

\bibitem[{{Nomoto}(1984)}]{Nomoto1984}
{Nomoto}, K. 1984, \apj, 277, 791

\bibitem[{{Nomoto}(1987)}]{Nomoto1987}
{Nomoto}, K. 1987, \apj, 322, 206

\bibitem[{{Nomoto} \& {Kondo}(1991)}]{Nomoto1991}
{Nomoto}, K. \& {Kondo}, Y. 1991, \apjl, 367, L19

\bibitem[{{Nomoto} {et~al.}(1984){Nomoto}, {Thielemann}, \&
  {Yokoi}}]{NomotoThielemann1984}
{Nomoto}, K., {Thielemann}, F.-K., \& {Yokoi}, K. 1984, \apj, 286, 644

\bibitem[{{Oda} {et~al.}(1994){Oda}, {Hino}, {Muto}, {Takahara}, \&
  {Sato}}]{ODA94}
{Oda}, T., {Hino}, M., {Muto}, K., {Takahara}, M., \& {Sato}, K. 1994, Atomic
  Data and Nuclear Data Tables, 56, 231

\bibitem[{{Ohlmann} {et~al.}(2014){Ohlmann}, {Kromer}, {Fink}, {Pakmor},
  {Seitenzahl}, {Sim}, \& {R{\"o}pke}}]{Ohlmann2014}
{Ohlmann}, S.~T., {Kromer}, M., {Fink}, M., {et~al.} 2014, \aap, 572, A57

\bibitem[{{Pakmor} {et~al.}(2012){Pakmor}, {Edelmann}, {R{\"o}pke}, \&
  {Hillebrandt}}]{Pakmor2012}
{Pakmor}, R., {Edelmann}, P., {R{\"o}pke}, F.~K., \& {Hillebrandt}, W. 2012,
  \mnras, 424, 2222

\bibitem[{{Paxton} {et~al.}(2011){Paxton}, {Bildsten}, {Dotter}, {Herwig},
  {Lesaffre}, \& {Timmes}}]{MESA2011}
{Paxton}, B., {Bildsten}, L., {Dotter}, A., {et~al.} 2011, \apjs, 192, 3

\bibitem[{{Paxton} {et~al.}(2013){Paxton}, {Cantiello}, {Arras}, {Bildsten},
  {Brown}, {Dotter}, {Mankovich}, {Montgomery}, {Stello}, {Timmes}, \&
  {Townsend}}]{MESA2013}
{Paxton}, B., {Cantiello}, M., {Arras}, P., {et~al.} 2013, \apjs, 208, 4

\bibitem[{{Paxton} {et~al.}(2015){Paxton}, {Marchant}, {Schwab}, {Bauer},
  {Bildsten}, {Cantiello}, {Dessart}, {Farmer}, {Hu}, {Langer}, {Townsend},
  {Townsley}, \& {Timmes}}]{MESA2015}
{Paxton}, B., {Marchant}, P., {Schwab}, J., {et~al.} 2015, \apjs, 220, 15

\bibitem[{{Poelarends} {et~al.}(2008){Poelarends}, {Herwig}, {Langer}, \&
  {Heger}}]{Poelarends2008}
{Poelarends}, A.~J.~T., {Herwig}, F., {Langer}, N., \& {Heger}, A. 2008, \apj,
  675, 614

\bibitem[{{Potekhin} \& {Chabrier}(2000)}]{Potekhin2000}
{Potekhin}, A.~Y. \& {Chabrier}, G. 2000, \pre, 62, 8554

\bibitem[{{Reinecke} {et~al.}(2002){Reinecke}, {Hillebrandt}, \&
  {Niemeyer}}]{Reinecke2002}
{Reinecke}, M., {Hillebrandt}, W., \& {Niemeyer}, J.~C. 2002, \aap, 391, 1167

\bibitem[{{Reinecke} {et~al.}(1999){Reinecke}, {Hillebrandt}, {Niemeyer},
  {Klein}, \& {Gr{\"o}bl}}]{Reinecke1999}
{Reinecke}, M., {Hillebrandt}, W., {Niemeyer}, J.~C., {Klein}, R., \&
  {Gr{\"o}bl}, A. 1999, \aap, 347, 724

\bibitem[{{R{\"o}pke} \& {Hillebrandt}(2005)}]{Roepke2005}
{R{\"o}pke}, F.~K. \& {Hillebrandt}, W. 2005, \aap, 431, 635

\bibitem[{{R{\"o}pke} {et~al.}(2006){R{\"o}pke}, {Hillebrandt}, {Niemeyer}, \&
  {Woosley}}]{Roepke2006}
{R{\"o}pke}, F.~K., {Hillebrandt}, W., {Niemeyer}, J.~C., \& {Woosley}, S.~E.
  2006, \aap, 448, 1

\bibitem[{{R{\"o}pke} \& {Schmidt}(2009)}]{Roepke2009}
{R{\"o}pke}, F.~K. \& {Schmidt}, W. 2009, in Interdisciplinary Aspects of
  Turbulence, ed. W.~{Hillebrandt} \& F.~{Kupka}, Lecture Notes in Physics
  (Berlin: Springer-Verlag), 255--289

\bibitem[{{R{\"o}pke} {et~al.}(2007){R{\"o}pke}, {Woosley}, \&
  {Hillebrandt}}]{Roepke2007}
{R{\"o}pke}, F.~K., {Woosley}, S.~E., \& {Hillebrandt}, W. 2007, \apj, 660,
  1344

\bibitem[{{Sana} {et~al.}(2012){Sana}, {de Mink}, {de Koter}, {Langer},
  {Evans}, {Gieles}, {Gosset}, {Izzard}, {Le Bouquin}, \&
  {Schneider}}]{Sana2012}
{Sana}, H., {de Mink}, S.~E., {de Koter}, A., {et~al.} 2012, Science, 337, 444

\bibitem[{{Schmidt} {et~al.}(2006){Schmidt}, {Niemeyer}, {Hillebrandt}, \&
  {R{\"o}pke}}]{Schmidt2006}
{Schmidt}, W., {Niemeyer}, J.~C., {Hillebrandt}, W., \& {R{\"o}pke}, F.~K.
  2006, \aap, 450, 283

\bibitem[{{Schwab} {et~al.}(2015){Schwab}, {Quataert}, \&
  {Bildsten}}]{Schwab2015}
{Schwab}, J., {Quataert}, E., \& {Bildsten}, L. 2015, \mnras, 453, 1910

\bibitem[{{Seitenzahl} {et~al.}(2011){Seitenzahl}, {Ciaraldi-Schoolmann}, \&
  {R{\"o}pke}}]{Seitenzahl2011}
{Seitenzahl}, I.~R., {Ciaraldi-Schoolmann}, F., \& {R{\"o}pke}, F.~K. 2011,
  \mnras, 414, 2709

\bibitem[{{Seitenzahl} {et~al.}(2013){Seitenzahl}, {Ciaraldi-Schoolmann},
  {R{\"o}pke}, {Fink}, {Hillebrandt}, {Kromer}, {Pakmor}, {Ruiter}, {Sim}, \&
  {Taubenberger}}]{Seitenzahl2013}
{Seitenzahl}, I.~R., {Ciaraldi-Schoolmann}, F., {R{\"o}pke}, F.~K., {et~al.}
  2013, \mnras, 429, 1156

\bibitem[{{Seitenzahl} {et~al.}(2015){Seitenzahl}, {Herzog}, {Ruiter},
  {Marquardt}, {Ohlmann}, \& {R{\"o}pke}}]{Seitenzahl2015}
{Seitenzahl}, I.~R., {Herzog}, M., {Ruiter}, A.~J., {et~al.} 2015, \prd, 92,
  124013

\bibitem[{{Seitenzahl} {et~al.}(2009){Seitenzahl}, {Townsley}, {Peng}, \&
  {Truran}}]{Seitenzahl2009}
{Seitenzahl}, I.~R., {Townsley}, D.~M., {Peng}, F., \& {Truran}, J.~W. 2009,
  Atomic Data and Nuclear Data Tables, 95, 96

\bibitem[{{Shibahashi} \& {Osaki}(1976)}]{Shibahashi1976}
{Shibahashi}, H. \& {Osaki}, Y. 1976, \pasj, 28, 199

\bibitem[{{Siess}(2007)}]{Siess2007}
{Siess}, L. 2007, \aap, 476, 893

\bibitem[{{Siess}(2010)}]{Siess2010}
{Siess}, L. 2010, \aap, 512, A10+

\bibitem[{{Smith}(2013)}]{Smith2013crab}
{Smith}, N. 2013, \mnras, 434, 102

\bibitem[{{Sullivan} {et~al.}(2016){Sullivan}, {O'Connor}, {Zegers}, {Grubb},
  \& {Austin}}]{Sullivan2016}
{Sullivan}, C., {O'Connor}, E., {Zegers}, R.~G.~T., {Grubb}, T., \& {Austin},
  S.~M. 2016, \apj, 816, 44

\bibitem[{{Takahara} {et~al.}(1989){Takahara}, {Hino}, {Oda}, {Muto},
  {Wolters}, {Glaudemans}, \& {Sato}}]{Takahara89}
{Takahara}, M., {Hino}, M., {Oda}, T., {et~al.} 1989, Nuclear Physics A, 504,
  167

\bibitem[{{Takahashi} {et~al.}(2013){Takahashi}, {Yoshida}, \&
  {Umeda}}]{Takahashi2013}
{Takahashi}, K., {Yoshida}, T., \& {Umeda}, H. 2013, \apj, 771, 28

\bibitem[{{Tauris} {et~al.}(2013){Tauris}, {Langer}, {Moriya}, {Podsiadlowski},
  {Yoon}, \& {Blinnikov}}]{Tauris2013}
{Tauris}, T.~M., {Langer}, N., {Moriya}, T.~J., {et~al.} 2013, \apjl, 778, L23

\bibitem[{{Timmes} \& {Arnett}(1999)}]{Timmes1999}
{Timmes}, F.~X. \& {Arnett}, D. 1999, \apjs, 125, 277

\bibitem[{{Timmes} \& {Woosley}(1992)}]{Timmes1992}
{Timmes}, F.~X. \& {Woosley}, S.~E. 1992, \apj, 396, 649

\bibitem[{{Toki} {et~al.}(2013){Toki}, {Suzuki}, {Nomoto}, {Jones}, \&
  {Hirschi}}]{Toki2013}
{Toki}, H., {Suzuki}, T., {Nomoto}, K., {Jones}, S., \& {Hirschi}, R. 2013,
  \prc, 88, 015806

\bibitem[{{Tominaga} {et~al.}(2013){Tominaga}, {Blinnikov}, \&
  {Nomoto}}]{Tominaga2013}
{Tominaga}, N., {Blinnikov}, S.~I., \& {Nomoto}, K. 2013, \apjl, 771, L12

\bibitem[{{Townsley} {et~al.}(2007){Townsley}, {Calder}, {Asida}, {Seitenzahl},
  {Peng}, {Vladimirova}, {Lamb}, \& {Truran}}]{Townsley2007}
{Townsley}, D.~M., {Calder}, A.~C., {Asida}, S.~M., {et~al.} 2007, \apj, 668,
  1118

\bibitem[{{Unno}(1967)}]{Unno1967}
{Unno}, W. 1967, \pasj, 19, 140

\bibitem[{{Woosley}(1986)}]{Woosley1986}
{Woosley}, S.~E. 1986, in Saas-Fee Advanced Course 16: Nucleosynthesis and
  Chemical Evolution, ed. J.~{Audouze}, C.~{Chiosi}, \& S.~E. {Woosley}, 1

\bibitem[{{Woosley} \& {Heger}(2015)}]{Woosley2015}
{Woosley}, S.~E. \& {Heger}, A. 2015, \apj, 810, 34

\bibitem[{{Zingale} \& {Dursi}(2007)}]{Zingale2007}
{Zingale}, M. \& {Dursi}, L.~J. 2007, \apj, 656, 333

\end{thebibliography}

\end{document}